\documentclass[prc,twocolumn,showpacs,floatfix,nofootinbib,preprintnumbers,superscriptaddress,amsmath,amssymb]{revtex4}

\usepackage{amsmath}           
\usepackage{amsfonts,mathrsfs}
\usepackage{graphicx}          
\usepackage{dcolumn}           

\begin{document}

\bibliographystyle{apsrev}

\preprint{\fbox{\sc version of \today}}

\title{One-quasiparticle States in the Nuclear Energy Density Functional Theory}

\author{N. Schunck}
\affiliation{Department of Physics and Astronomy, University of Tennessee, Knoxville, Tennessee 37996, USA}
\affiliation{Physics Division, Oak Ridge National Laboratory, P.O. Box 2008, Oak Ridge, Tennessee 37831, USA}

\author{J. Dobaczewski}
\affiliation{Institute of Theoretical Physics, Warsaw University, ul. Ho\'{z}a 69, PL-00681, Warsaw, Poland}
\affiliation{Department of Physics, P.O. Box 35 (YFL), FI-40014 University of Jyv\"{a}skyl\"{a}, Finland}
\affiliation{Joint Institute for Heavy Ion Research, Oak Ridge National Laboratory, P.O. Box 2008, Oak Ridge, TN 37831, USA}

\author{J. McDonnell}
\affiliation{Department of Physics and Astronomy, University of Tennessee, Knoxville, Tennessee 37996, USA}
\affiliation{Physics Division, Oak Ridge National Laboratory, P.O. Box 2008, Oak Ridge, Tennessee 37831, USA}

\author{J. Mor\'e}
\affiliation{Mathematics and Computer Science Division, Argonne National Laboratory, Argonne, IL 60439-4844, USA}

\author{W. Nazarewicz}
\affiliation{Department of Physics and Astronomy, University of Tennessee, Knoxville, Tennessee 37996, USA}
\affiliation{Physics Division, Oak Ridge National Laboratory, P.O. Box 2008, Oak Ridge, Tennessee 37831, USA}
\affiliation{Institute of Theoretical Physics, Warsaw University, ul. Ho\'{z}a 69, PL-00681, Warsaw, Poland}

\author{J. Sarich}
\affiliation{Mathematics and Computer Science Division, Argonne National Laboratory, Argonne, IL 60439-4844, USA}

\author{M. V.~Stoitsov}
\affiliation{Department of Physics and Astronomy, University of Tennessee, Knoxville, Tennessee 37996, USA}
\affiliation{Physics Division, Oak Ridge National Laboratory, P.O. Box 2008, Oak Ridge, Tennessee 37831, USA}
\affiliation{Institute of Nuclear Research and Nuclear Energy, Bulgarian Academy of Sciences, Sofia, Bulgaria}

\date{\today}

\newcommand{\BdG}{\textsc{b}{\footnotesize d}\textsc{g}}
\newcommand{\PWscf}{\textsc{pw}scf}
\newcommand{\DFT}{\textsc{dft}}
\newcommand{\RMF}{\textsc{rmf}}
\newcommand{\SLDA}{\textsc{slda}}
\newcommand{\HO}{\textsc{ho}}
\newcommand{\THO}{\textsc{tho}}
\newcommand{\HF}{\textsc{hf}}
\newcommand{\BCS}{\textsc{bcs}}
\newcommand{\LN}{\textsc{ln}}
\newcommand{\EFA}{\textsc{efa}}
\newcommand{\HFB}{\textsc{hfb}}
\newcommand{\HFODD}{\textsc{hfodd}}
\newcommand{\HFBTHO}{\textsc{hfbtho}}
\newcommand{\EDF}{\textsc{edf}}
\newcommand{\RPA}{\textsc{rpa}}
\newcommand{\CC}{\textsc{cc}}
\newcommand{\CCSD}{\textsc{ccsd}}
\newcommand{\BLAS}{\textsc{blas}}
\newcommand{\LAPACK}{\textsc{lapack}}
\newcommand{\ATLAS}{\textsc{atlas}}
\newcommand{\GNU}{\textsc{gnu}}
\newcommand{\CECO}{\textsc{ceco}}
\newcommand{\exclude}[1]{}
\newcommand{\mat}[1]{\mathbf{#1}}
\newcommand{\ket}[1]{|#1\rangle}
\newcommand{\bra}[1]{\langle#1|}
\newcommand{\norm}[1]{\lVert{#1}\rVert}
\newcommand{\braket}[1]{\mathinner{\langle{#1}\rangle}}{\catcode`\|=\active
  \gdef\Braket#1{\left<\mathcode`\|"8000\let|\bravert {#1}\right>}}
\newcommand{\bravert}{\egroup\,\vrule\,\bgroup}

\newcommand{\gras}[1]{\boldsymbol{#1}}
\newcommand{\capital}[1]{\mathscr{#1}}

\newcommand{\ba}{\begin{array}}
\newcommand{\ea}{\end{array}}

\newcommand{\disregard}[1]{}

\newcommand{\ali}{ali}

\begin{abstract}
We study one-quasiproton excitations in the rare-earth region in the
framework of the nuclear Density Functional Theory in the
Skyrme-Hartree-Fock-Bogoliubov variant. The blocking prescription is
implemented exactly, with the time-odd mean field fully taken into
account. The equal filling approximation is compared with the exact
blocking procedure. We show that both procedures are strictly
equivalent when the time-odd channel is neglected, and discuss how
nuclear alignment properties affect  the time-odd fields. The impact
of time-odd fields on calculated one-quasiproton bandhead energies is
found to be rather small, of the order of  100--200\,keV; hence, the
equal filling approximation is sufficiently precise for most practical
applications. The triaxial polarization of the core induced by the
odd particle is studied. We also briefly discuss the occurrence of
finite-size spin instabilities that are present in calculations for
odd-mass nuclei when certain Skyrme functionals are employed.
\end{abstract}

\pacs{21.60.Jz, 21.10.Pc, 21.30.Fe, 27.70.+q}

\maketitle

\section{Introduction}
\label{Sec-Introduction}

The nuclear Density Functional Theory ({\DFT})
\cite{(Pet91),[Ben03],[Lal04]} plays a central role in the quest for
a microscopic and quantitative description of atomic nuclei. The
energy functionals related to effective two-body density-dependent
interactions are the main building blocks of the mean-field theory of
the nucleus wherein the self-consistency is imposed through the
Hartree-Fock-Bogoliubov ({\HFB}) formalism. This framework has
provided a consistent description of a broad range of phenomena
ranging from nuclear masses to collective excitations. Over the last
few years, however, with the influx of high-quality experimental data on
exotic nuclei, it has become evident that the standard local
functionals (e.g., extended Skyrme functionals) are too restrictive
when one is aiming at detailed quantitative description and
extrapolability \cite{[Sto06],[Ber05],[Zal08],[Kor08]}. Consequently,
various strategies have been devised to develop realistic nuclear
energy density functionals ({\EDF}) \cite{[Ber07a]}. These include:
(i) the use of the density matrix expansion technique
\cite{[Neg72],[Bog09]} to relate the functional to low-momentum
interactions; (ii) extending {\EDF} by adding higher-order terms in
the local densities \cite{[Car08]}; and (iii) improving spin and
isospin properties \cite{[Les07],[Sat08],[Sat09],[Mar09]}. In any case,
regardless of the strategy, the fine-tuning of the coupling constants
of the functional to a suitably chosen  set of experimental data is
necessary to provide quality description \cite{[Gor09]}.

When aiming at spectroscopic-quality functionals \cite{[Zal08]},
the data coming from odd-mass nuclei are crucial: the energies,
angular momenta, and parities of one-quasiparticle excitations
provide us with basic knowledge about the underlying shell structure.
Moreover, binding energies of odd-$A$ systems  are instrumental
for determining the magnitude of collective effects such as pairing.
Theoretically, however, since nuclei with an odd number of particles
have non-zero angular momentum ($J>0$), i.e., they are spin-polarized,
their treatment is considerably more involved as compared to the
$J^\pi$=$0^+$ ground-state (g.s.) configurations of doubly-even nuclei.

Mathematically, the local {\EDF} is a time-even scalar constructed
from various local densities and currents related to particle and
pairing density distributions \cite{[Eng75],[Per04]}. The resulting
mean field contains both time-even and  time-odd terms. While the
time-odd fields automatically vanish in the ground state of
doubly-even nuclei, they are non-zero in $J>0$ configurations in
which time-reversal symmetry is internally broken \cite{[Ben03],[Lal04]}.
The time-odd fields have been investigated in the context of high-spin
states \cite{[Pos85wf],[Che92],[Dob95e],[Afa00]}, Gamow-Teller
excitations \cite{[Ben02]}, single-particle (s.p.) spectra
\cite{[Rut98],[Zal08],[Sat08]}, and collective dynamics
\cite{[Bar78w],[Dob81],[Mar06],[Hin06]}. The general consensus is
that they can appreciably impact the nuclear collective motion. On
the other hand, our knowledge of the coupling constants
characterizing individual time-odd fields is fairly limited, and the

impact of those terms on nuclear ground-states still needs to be
assessed. Conversely, one can ask whether experimental data on
nuclear ground states can help constrain the time-odd fields of the
nuclear {\EDF}.

There have been very few systematic theoretical studies of
one-quasiparticle states along isotopic or isotonic chains.
Regional systematics of one-quasiparticle excitations, and their
consequences on various observables in spherical
and deformed nuclei, can be found in, e.g.,
Refs.~\cite{[Ogl71w],[Naz90],[Cwi91w],[Cwi94a],[Par05]}
(macroscopic-microscopic approach) and
Refs.~\cite{[Rut98],[Rut99fw],[Cwi99],[Afa03],[Sat08],[Afa10]}
(nuclear DFT). The only global DFT study of ground state spin and
parity for odd-mass nuclei is that of Bonneau {\it et al.}
\cite{[Bon07]}. It is to be noted, however, that most of these
studies were restricted in one way or another, e.g., by assuming
axial symmetry, neglecting the time-odd fields, or doing an
approximate treatment  of blocking. The results of
Refs.~\cite{[Bon07],[Zal08],[Kor08]} clearly indicate that the
currently used nuclear density  functionals give a rather poor
description of s.p.\ states, so it is imperative to evaluate the
magnitude of the effects due to theoretical limitations and
approximations.

The goal of this study is to review the description of odd-mass
nuclei in the framework of the nuclear {\DFT} and assess the
magnitude of time-odd polarizations through large-scale surveys. We
compare various treatments of blocking, associated approximations,
and resulting uncertainties. We discuss the choice of the orientation
of the alignment vector, which is important for maintaining s.p.\
characteristics during the blocking procedure. We also assess the
impact of the time-odd fields on binding energies of one-quasiparticle
states and estimate the polarization due to the axial symmetry
breaking in certain orbits.

This paper is organized as follows. Section~\ref{Sec-DFT} summarizes
the main features of the nuclear Skyrme-{\DFT}. We pay special
attention to the treatment of odd-mass nuclei through the so-called
blocking approximation and the Equal Filling Approximation ({\EFA}).
In Sec.~\ref{Sec-Method} we present the details of the calculations
and discuss various optimization techniques that enable  large-scale
calculations for odd-mass nuclei. The results are presented in
Sec.~\ref{Sec-Results}. We first compare the {\EFA} approximation
with the exact blocking prescription. We estimate the effect of the
time-odd fields on one-quasiparticle states in the rare-earth region
and make selected comparisons with experiment. We also comment on the
finite-size instabilities related to certain energy functionals that
show up when studying polarized systems. Finally, the conclusions are
contained in Sec.~\ref{Sec-Conclusions}.


\section{DFT Treatment of One-Quasiparticle states}
\label{Sec-DFT}

The nuclear {\DFT} in a Skyrme variant has been presented in great
detail in a number of articles \cite{[Ben03],[Per04],[Sto07b]}. In
the following we recall only the salient features of the theory that
will be needed in this study.


\subsection{Representations of the Density Matrix}
\label{Sec-DFT-Representations}

The cornerstone of the nuclear {\DFT} is the general one-body density
operator $\hat{\rho}$. Two representations of the density matrix are
often considered. In the coordinate representation, the s.p.\ space is
spanned by the continuous basis of states
$|\gras{r}\sigma\rangle = |\gras{r}\rangle\otimes|\sigma\rangle$
\cite{[Ben03],[Bul80],[Dob84],[Dob96]}. In the configuration
representation, a basis of discrete states $|n\rangle$ is introduced,
where $n$ stands for all the s.p.\ quantum numbers. The choice of one
particular representation depends on the context.

If $|\Phi\rangle$ is the many-body  state, the non-local density
matrix in coordinate representation reads:
\begin{equation}
\rho(\gras{r}\sigma,\gras{r'}\sigma')
= \braket{\Phi | c_{\gras{r'}\sigma'}^{\dagger}c_{\gras{r}\sigma} | \Phi},
\label{density-matrix-R}
\end{equation}
where $c^{\dagger}_{\gras{r}\sigma}$ is a fermionic field operator
creating a particle at position $\gras{r}$ with spin projection
$\sigma$ and $c_{\gras{r}\sigma}$ is the corresponding annihilation
operator. The field operators can be expressed in terms of the
standard fermionic creation and annihilation operators
$c_{n}^{\dagger}$ and $c_{n}$ associated with the basis $|n\rangle$
\cite{(Bla86),(Rin80)}:
\begin{subequations}
\begin{eqnarray}
c_{\gras{r}\sigma}^{\dagger} = \displaystyle\sum_{n} \phi_{n}^{*}(\gras{r}\sigma) c_{n}^{\dagger}, \label{CtoR-1}\medskip\\
c_{\gras{r}\sigma}           = \displaystyle\sum_{n} \phi_{n}    (\gras{r}\sigma) c_{n}. \label{CtoR-2}
\end{eqnarray}
\end{subequations}
Note that in this expression, $\phi_{n}(\gras{r}\sigma)$ and
$\phi_{n}^{*}(\gras{r}\sigma)$ are matrix elements of the basis
transformation $|\gras{r}\sigma\rangle \leftrightarrow |n\rangle$:
$\phi_{n}(\gras{r}\sigma) = \langle\gras{r}\sigma| n\rangle$ and
$\phi_{n}^{*}(\gras{r}\sigma) = \langle n| \gras{r}\sigma\rangle$.
They are therefore complex s.p.\ wave functions dependent on the
position vector $\gras{r}$ and spin coordinate $\sigma$. The inverse
relations are:
\begin{subequations}
\begin{eqnarray}
c_{n}^{\dagger} = \displaystyle\int d^{3}\gras{r}\sum_{\sigma} \phi_{n}(\gras{r}\sigma) c_{\gras{r}\sigma}^{\dagger}, \label{RtoC-1}\medskip\\
c_{n}           = \displaystyle\int d^{3}\gras{r}\sum_{\sigma} \phi_{n}^{*}(\gras{r}\sigma) c_{\gras{r}\sigma}. \label{RtoC-2}
\end{eqnarray}
\end{subequations}
For complete bases, relations (\ref{CtoR-1}-\ref{CtoR-2}) and
(\ref{RtoC-1}-\ref{RtoC-2}) allow us to express the relations
between the two representations, $\rho(\gras{r}\sigma,\gras{r'}\sigma')$
and $\rho_{mn}$ of the density matrix.

The density matrix (\ref{density-matrix-R}) can be regarded as the
matrix element of an operator
$\hat{\rho}(\gras{r}\sigma,\gras{r'}\sigma')$ acting in the spin
space. Any such operator can be expressed in terms of  the Pauli
matrices $\gras{\sigma}$ and the identity matrix. This  leads to a
spin-scalar  $\hat{\rho}(\gras{r},\gras{r'})$ and a spin-vector field
$\hat{\gras{s}}(\gras{r},\gras{r'})$. These two fields are the
fundamental building blocks of the nuclear {\DFT}.


\subsection{Skyrme Energy Functional}
\label{Sec-DFT-EDF}

The contribution to the total energy of the system coming from the
Skyrme interaction reads:
\begin{equation}
E^{\text{Skyrme}} = \sum_{t=0,1}\int d^{3}\gras{r} \left\{ {\cal H}_{t}^{\text{(even)}}(\gras{r}) + {\cal H}_{t}^{\text{(odd)}}(\gras{r}) \right\},
\label{Skyrme_energy}
\end{equation}
where $t$=0 and $t$=1 corresponds to isoscalar and isovector components,
respectively. In this paper, we do not consider proton-neutron mixing
\cite{[Per04]}. Using the standard notation for the local densities and
currents \cite{[Eng75],[Per04]}, the part of the energy density that
depends  on  time-even fields can be written as:
\begin{multline}
{\cal H}_{t}^{\text{(even)}}(\gras{r}) =
C_{t}^{\rho}\rho_{t}^{2} +
C_{t}^{\Delta\rho}\rho_{t}\Delta\rho_{t} +
C_{t}^{\tau}\rho_{t}\tau_{t}  \\
+
C_{t}^{J} \overleftrightarrow{J}_{t}^{2}
+
C_{t}^{\nabla J} \rho_{t}\gras{\nabla}\cdot\gras{J}_{t},
\label{Heven}
\end{multline}
while the part depending on the time-odd fields is:
\begin{multline}
{\cal H}_{t}^{\text{(odd)}}(\gras{r}) =
C_{t}^{s}\gras{s}_{t}^{2} +
C_{t}^{\Delta s}\gras{s}_{t}\cdot\Delta\gras{s}_{t}
+
C_{t}^{T}\gras{s}_{t}\cdot\gras{T}_{t}  \\
+
C_{t}^{j} \gras{j}_{t}^{2}
+
C_{t}^{\nabla j} \gras{s}_{t}\cdot\left(\gras{\nabla}\wedge\gras{j}_{t}\right).
+
C_{t}^{F}\gras{s}_{t}\cdot\gras{F}_{t} ,
\label{Hodd}
\end{multline}
All densities and currents  entering Eqs.~(\ref{Heven}) and
(\ref{Hodd}) can be related to the particle density $\rho(\gras{r},\gras{r'})$,
spin density $\gras{s}(\gras{r},\gras{r'})$, and their derivatives
\cite{[Eng75],[Per04]}. In the present work, we do not
consider tensor interactions and therefore we set $C_{t}^{F}=0$.

Below, we discuss several versions of the functional, depending on
how the time-odd coupling constants are determined:
\begin{itemize}
\item Native version, which corresponds to all time-odd
coupling constants being determined by the underlying Skyrme interaction
\cite{[Per04]}.
\item Gauge version, which corresponds to the subset of time-odd
coupling constants being determined through the gauge-invariance
conditions \cite{[Dob95e],[Per04]}, namely,
$C_{t}^{j}        = - C_{t}^{\tau}    $,
$C_{t}^{T}        = - C_{t}^{J}       $, and
$C_{t}^{\nabla j} =   C_{t}^{\nabla J}$, and all other time-odd
coupling constants set to zero.
\item Landau version, which is based on the gauge version where
the subset of time-odd coupling constants $C_{t}^{s}$ and
$C_{t}^{T}$ are reset through the Landau parameters \cite{[Ben02]}:
\begin{equation}
\begin{array}{ccl}
g_{0}  & = &   N_{0}\left(2C_{0}^{s} + 2C_{0}^{T}\beta\rho_{0}^{2/3}\right), \\
g_{1}  & = & -2N_{0}C_{0}^{T}\beta\rho_{0}^{2/3}, \\
g'_{0} & = &   N_{0}\left(2C_{1}^{s} + 2C_{1}^{T}\beta\rho_{0}^{2/3}\right),\\
g'_{1} & = & -2N_{0}C_{1}^{T}\beta\rho_{0}^{2/3},
\end{array}
\end{equation}
where $\beta = (3\pi^{2}/2)^{2/3}$, $1/N_{0} = \pi^{2}\hbar^{2}/2m^{\star}k_{F}$
and, additionally, $C_{t}^{\Delta s} = 0$ for $t = 0,1$. Since the
Landau prescription only sets $C_{t}^{T}$, the gauge condition is
broken since $C_{t}^{T} \neq - C_{t}^{J}$ anymore.
\item Time-even version, in which all time-odd coupling constants in
Eq.~(\ref{Hodd}) are set equal to zero.
\end{itemize}


\subsection{HFB Method}
\label{Sec-DFT-Pairing}

In the {\HFB} theory, pairing correlations enter through the pairing
tensor $\kappa$ defined in coordinate representation as:
\begin{equation}
\kappa(\gras{r}\sigma,\gras{r'}\sigma')
=
\braket{\Phi | c_{\gras{r'}\sigma'}c_{\gras{r}\sigma} | \Phi},
\end{equation}
(From a practical point of view, it is sometimes more advantageous to
use the pairing density $\tilde{\rho}$ \cite{[Dob84],[Dob96]}.)

The starting point of the {\HFB} theory is to assume that the
ground-state of an even-even nucleus is a vacuum for quasiparticle
operators $(\beta_{\nu}, \beta_{\nu}^{\dagger})$. The latter are
obtained from single-particle operators $(c_{n}, c_{n}^{\dagger})$
associated with the single-particle basis states $|n\rangle$ by the
Bogoliubov transformation:
\begin{subequations}
\label{Bogo}
\begin{eqnarray}
\beta_{\nu} = \sum_{n} U^{*}_{n\nu}c_{n} + V^{*}_{n\nu}c^{\dagger}_{n}, \\
\beta^{\dagger}_{\nu} = \sum_{n} V_{n\nu}c_{n} + U_{n\nu}c^{\dagger}_{n}.
\end{eqnarray}
\end{subequations}
The matrices $U$ and $V$ are obtained from  the {\HFB} equations:
\begin{equation}
\left(
\begin{array}{cc}
\hat{h}-\lambda & \hat\Delta \\
-\hat\Delta^{*} & -\hat{h}^{*}+\lambda
\end{array}
\right)
\left( \begin{array}{c}
U\\
V
\end{array}
\right)
=
E
\left( \begin{array}{c}
U\\
V
\end{array}
\right),
\label{HFB_basis}
\end{equation}
where  $\lambda$ is the chemical potential, $\hat{h}$ is the Hartree-Fock ({\HF})
potential and $\hat\Delta$ the pairing potential. (From a practical point of view,
it is sometimes more advantageous to use the pairing potential $\hat{\tilde{h}}$
\cite{[Dob84],[Dob96]}.) The form of the {\HFB} equations in coordinate space can
be found in Refs.~\cite{[Dob84],[Dob96]}.

The density matrix and pairing tensor can be written as:
\begin{subequations}
\begin{eqnarray}
\rho_{mn}   = \left( V^{*}V^{T} \right)_{mn}, \label{densities-C-1}\medskip\\
\kappa_{mn} = \left( V^{*}U^{T} \right)_{mn}. \label{densities-C-2}
\end{eqnarray}
\end{subequations}
The coordinate representation of the Bogoliubov transformation,
\begin{subequations}
\begin{eqnarray}
\beta_{\nu}            = \int d^{3}\gras{r}\sum_{\sigma}
                        \left\{ U^{(\nu)*}(\gras{r}\sigma) c_{\gras{r}\sigma} +
                                V^{(\nu)*}(\gras{r}\sigma) c^{\dagger}_{\gras{r}\sigma}
                        \right\},\\
\beta^{\dagger}_{\nu}  = \int d^{3}\gras{r}\sum_{\sigma}
                        \left\{ V^{(\nu)}(\gras{r}\sigma) c_{\gras{r}\sigma} +
                                U^{(\nu)}(\gras{r}\sigma) c^{\dagger}_{\gras{r}\sigma}
                        \right\},
\end{eqnarray}
\end{subequations}
can be expressed through lower and upper components of the quasi-particle (q.p.)
wave functions:
\begin{subequations}
\begin{eqnarray}
V^{(\nu)}(\gras{r}\sigma) = \sum_{n} \phi_{n}^{*}(\gras{r}\sigma)V_{n\nu}, \label{QP-CtoR-1}\\
U^{(\nu)}(\gras{r}\sigma) = \sum_{n} \phi_{n}    (\gras{r}\sigma)U_{n\nu}.
\label{QP-CtoR-2}
\end{eqnarray}
\end{subequations}
Finally,  the density matrix and pairing tensor in coordinate space are:
\begin{subequations}
\begin{eqnarray}
\displaystyle\rho(\gras{r}\sigma,\gras{r'}\sigma') =
\sum_{0 \leq E_{\mu} \leq E_{\text{max}}} V^{(\mu)*}(\gras{r}\sigma) V^{(\mu)}(\gras{r'}\sigma'), \label{densities-R-1}\medskip\\
\displaystyle\kappa(\gras{r}\sigma,\gras{r'}\sigma') =
\sum_{0 \leq E_{\mu}  \leq E_{\text{max}}} V^{(\mu)*}(\gras{r}\sigma) U^{(\mu)}(\gras{r'}\sigma').\label{densities-R-2}
\end{eqnarray}
\end{subequations}
It is assumed here that the q.p.\ continuum with $E > -\lambda$ has
been discretized and all q.p.\ states with energy lower than some
cut-off energy $E_{cut}$ are retained (see discussion in
Ref.~\cite{[Dob84]}).


\subsection{The Blocking Prescription and the Equal Filling Approximation}
\label{Sec-DFT-Blocking}

In the {\HFB} theory, the ground-state of an odd nucleus is a one
quasiparticle excitation $\beta_{\mu_0}^{\dagger}$ with respect to the
 q.p.\ vacuum. In the configuration representation, the
corresponding density matrix and pairing tensor are
\cite{[Ban74],[Fae80],[Ber09],[Per08]}:
\begin{subequations}
\begin{eqnarray}
\rho_{mn}^{B,\mu_0}   = \left( V^{*}V^{T} \right)_{mn} + U_{m\mu_{0}}U^{*}_{n\mu_{0}} - V^{*}_{m\mu_{0}}V_{n\mu_{0}}, \label{exactBlocking-C-1} \\
\kappa_{mn}^{B,\mu_0} = \left( V^{*}U^{T} \right)_{mn} + U_{m\mu_{0}}V^{*}_{n\mu_{0}} - V^{*}_{m\mu_{0}}U_{n\mu_{0}}. \label{exactBlocking-C-2}
\end{eqnarray}
\end{subequations}
In practice, one must adopt a prescription to be able to determine,
at each iteration, the index $\mu_{0}$ of the quasiparticle state to
be blocked \cite{[Hee95]}. In the present study, this has been done
according to the recipe described in Ref.~\cite{[Dob09d]}. In the first
step, the mean field Hamiltonian $\hat{h}$ is diagonalized:
\begin{equation}
\hat{h} \varphi_{n} = e_n \varphi_{n}.
\label{equivalent-spectrum}
\end{equation}
Since in this work parity and $y$-signature are assumed to be
self-consistent symmetries, every s.p.\ level $e_n$ is  uniquely
identified by its position in a given parity and $y$-signature block.
This unique identification allows to pin down the configuration of
the blocking candidate $n_{0}$. To connect the s.p.\ state
$\varphi_{n_{0}}$  with a quasiparticle state to be blocked, we
calculate at each iteration the overlap between $\varphi_{n_{0}}$ and
both the upper component $U_{\mu}$ and the time-reversed lower
component $V_{\bar{\mu}}$ of  quasiparticle states around the Fermi
level \cite{[Dob09d]}. The largest overlap in this set defines the
index $\mu_{0}$ of the quasiparticle state to be blocked.  In the
beginning of the iterative process, s.p.\  states of a neighboring
even-even nucleus are taken.

Within the {\EFA}, the states $\mu_{0}$ and its time-reversal partner
$\bar{\mu}_{0}$ enter the density matrix and pairing tensor with the
same weights, which ensures time-reversal symmetry and thereby degeneracy
of $\mu_{0}$ and $\bar{\mu}_{0}$ \cite{[Per08]}:
\begin{subequations}
\begin{align}
\begin{split}
\rho_{mn}^{{\EFA},\mu_0} = \left( V^{*}V^{T} \right)_{mn} +
\frac{1}{2}\left(
U_{m\mu_{0}}U^{*}_{n\mu_{0}} - V^{*}_{m\mu_{0}}V_{n\mu_{0}} \right.\\
\left. +
U_{m\bar{\mu}_{0}}U^{*}_{n\bar{\mu}_{0}} - V^{*}_{m\bar{\mu}_{0}}V_{n\bar{\mu}_{0}}
\right), \label{EFA-1}
\end{split}\\
\begin{split}
\kappa_{mn}^{{\EFA},\mu_0} =
\left( V^{*}U^{T} \right)_{mn}
+ \frac{1}{2}
\left(
  U_{m\mu_{0}}      V^{*}_{n\mu_{0}}       - V^{*}_{m\mu_{0}}      U_{n\mu_{0}} \right.\\
\left.
+ U_{m\bar{\mu}_{0}}V^{*}_{n\bar{\mu}_{0}} - V^{*}_{m\bar{\mu}_{0}}U_{n\bar{\mu}_{0}}
\right). \label{EFA-2}
\end{split}
\end{align}
\end{subequations}
The  {\HFB} equations are then solved by replacing ($\rho^{B,\mu_0}$,
$\kappa^{B,\mu_0}$) with ($\rho^{\EFA,\mu_0}$, $\kappa^{\EFA,\mu_0}$).
For the justification of the {\EFA} ansatz by means of statistical
density operators and for detailed discussion of the procedure involved,
we refer the reader to Refs.~\cite{[Per08],[Per07]}.

In this work we point to another possible justification of the
{\EFA}. We first notice that the time-even parts of the blocked density
matrices given by Eqs.~(\ref{exactBlocking-C-1}) and (\ref{exactBlocking-C-2})
are identical to the time-even parts of the density matrices in {\EFA},
Eqs.~(\ref{EFA-1}) and (\ref{EFA-2}). Therefore, all time-even densities in
Eq.~(\ref{Heven}) are exactly the same in both variants. Consequently,
in the blocking and {\EFA} approximations, the time-even part of the
functional (Sec.~\ref{Sec-DFT-EDF}) yields exactly the same self-consistent
solution. This allows us to reinterpret {\EFA} density matrices as those
corresponding to the time-even functional in which the time-odd polarizations
exerted by the odd particle are dynamically switched off. Of course, the
blocking prescription and {\EFA} give exactly the same average values of
all time-even observables (e.g., radii and multipole moments) but
differ in the average values of time-odd observables  (e.g., spin
alignments and magnetic moments).


\subsection{Blocking, Alignments, and Symmetries}
\label{Sec-DFT-Symmetries}

Although for the functionals restricted to time-even fields (or within
{\EFA}) the time-reversed q.p\ states
$|\mu\rangle=\beta^{\dagger}_{\mu}|0\rangle$ and
$|\bar{\mu}\rangle=\beta^{\dagger}_{\bar{\mu}}|0\rangle$ are exactly
degenerate, this is not true any more in the general case. Here, the
blocking prescription does depend on which of these two states, or
which linear combination thereof, is used in
Eqs.~(\ref{exactBlocking-C-1}) and (\ref{exactBlocking-C-2}). In
order to discuss this point, we introduce here the notion of an
``{\ali}spin'', which pertains to the unitary mixing of states
$|\mu\rangle$ and $|\bar{\mu}\rangle$. This is in complete analogy
with the standard notion of the isospin, which involves the unitary
mixing of proton and neutron states.

An alivector $\mathcal{V}^{(\mu)}$ is defined as a set of two complex
numbers $a$ and $b$ ($|a|^{2} + |b|^{2} = 1$):
\begin{equation}
\mathcal{V}^{(\mu)} = \left( \begin{array}{c} a \\ b \end{array} \right),
\end{equation}
which corresponds to the linear combination of states $|\mu\rangle$ and
$|\bar{\mu}\rangle$: $|v_{\mu}\rangle = a|\mu\rangle + b|\bar{\mu}\rangle$.
Alivectors reside in SU(2) space; therefore the {\ali}rotation by an
angle $\gras{\phi}^{(\mu)}$ is defined as:
\begin{equation}
\label{alirot}
\mathcal{V}^{(\mu)'} =
\left(\ba{c}a' \\ b' \ea\right)
=e^{i\gras{\phi}^{(\mu)}\circ\gras{\sigma}^{(\mu)}}
\left(\ba{c}a \\ b \ea\right),
\end{equation}
where the {\ali}vectors of Pauli matrices are denoted by $\gras{\sigma}^{(\mu)}$,
and $\circ$ denotes the scalar product of {\ali}vectors. To recall that
the {\ali}rotation pertains to a single pair of states, we use superscripts
$(\mu)$ throughout.

The blocked density matrix Eq.~(\ref{exactBlocking-C-1}) corresponding
to the state $\mathcal{V}^{(\mu)}$ reads:
\begin{multline}
\rho_{mn}^{B,(a,b)} =
\rho_{mn}
- \left\{ |a|^{2}V_{n\mu}V_{m\mu}^{*}
+ |b|^{2}V_{n\bar{\mu}}V_{m\bar{\mu}}^{*} \right. \\
\left.
+ a^{*}b V_{n\bar{\mu}}V_{m\mu}^{*}
+ ab^{*}V_{n\mu}V_{m\bar{\mu}}^{*} \right\} \\
+ \left\{ |a|^{2}U_{n\mu}^{*}U_{m\mu}
+ |b|^{2}U_{n\bar{\mu}}^{*}U_{m\bar{\mu}} \right. \\
\left.
+ a^{*}b U_{n\bar{\mu}}^{*}U_{m\mu}
+ ab^{*}U_{n\mu}^{*}U_{m\bar{\mu}} \right\}.
\end{multline}
If time-reversal symmetry is conserved, the different blocks of the Bogoliubov matrices are related:
\begin{equation}
-V_{\bar{n}\mu}^{*} = V_{n\bar{\mu}}\ \ \ \text{and}\ \ \ V_{n\mu}^{*} = V_{\bar{n}\bar{\mu}},
\label{Teven-1}
\end{equation}
and $\rho_{mn}^{(a,b)} = \rho_{\bar{m}\bar{n}}^{(a,b) *}$. These
relations lead to:
\begin{equation}
\rho_{mn}^{(ab)} = \rho_{mn} - V_{n\mu}V_{m\mu}^{*} + U_{n\mu}^{*}U_{m\mu},
\end{equation}
or, equivalently
\begin{equation}
\rho_{mn}^{(ab)} =
\rho_{mn} - V_{n\bar{\mu}}V_{m\bar{\mu}}^{*} + U_{n\bar{\mu}}^{*}U_{m\bar{\mu}}.
\end{equation}
Therefore, in this limit, the exact blocking density matrix becomes independent
of the coefficients $(a,b)$ of the mixing, i.e., it is an {\ali}scalar. Since
\begin{equation}
\rho_{mn}^{EFA} = \frac{1}{2}\left( \rho_{mn}^{(1,0)} + \rho_{mn}^{(0,1)} \right),
\end{equation}
the {\EFA} density matrix also coincides with the exact blocking density matrix;
hence, it is  an {\ali}scalar as well.

In the general case where time-reversal symmetry is not dynamically conserved,
however, the blocking density matrix is not {\ali}scalar and the energy of the
system may change as a function of the  mixing coefficients $(a,b)$. To
analyze what are the consequences of blocking different {\ali}rotated states
$\mathcal{V}^{(\mu_{0})'}$, we introduce the (real) alignment vector
$\gras{J}^{(\mu)}=\langle\mu|\hat{\gras{J}}|\mu\rangle
=-\langle\bar{\mu}|\hat{\gras{J}}|\bar{\mu}\rangle$ and the (complex)
decoupling vector
$\gras{D}^{(\mu)}=\langle\mu|\hat{\gras{J}}|\bar{\mu}\rangle
=\langle\bar{\mu}|\hat{\gras{J}}|\mu\rangle^*$ \cite{[Olb06]}.
Together, they form the matrix elements of the alignment vector-{\ali}vector
$\hat{\mathcal{J}}^{(\mu)}$:
\begin{equation}
\hat{\mathcal{J}}^{(\mu)} =
\left(
\begin{array}{cc}
\gras{J}^{(\mu)}     &  \gras{D}^{(\mu)} \\
\gras{D}^{(\mu)*} & -\gras{J}^{(\mu)}
\end{array}
\right).
\end{equation}
Expanding this operator (acting on SU(2) {\ali}states) in the basis of Pauli
matrices, we find:
\begin{subequations}
\begin{align}
\hat{\mathcal{J}}^\mu_1 &= +\Re\gras{D}^\mu  ,
\label{ali-1}
\\
\hat{\mathcal{J}}^\mu_2 &= -\Im\gras{D}^\mu  ,
\label{ali-2}
\\
\hat{\mathcal{J}}^\mu_3 &= +\gras{J}^\mu  ,
\label{ali-3}
\end{align}
\end{subequations}
where indices $k=1,2,3$ enumerate the components of {\ali}vectors. From
these considerations it follows that the alignment vector-{\ali}vector
$\hat{\mathcal{J}}^{\mu'}$, which corresponds to the {\ali}rotated pair
(\ref{alirot}), is obtained by:
\begin{equation}
\left(\ba{l} \hat{\mathcal{J}}^{\mu'}_1 \\
             \hat{\mathcal{J}}^{\mu'}_2 \\
             \hat{\mathcal{J}}^{\mu'}_3 \ea\right)
=\exp(i\vec{\phi}_\mu\circ\vec{S}_\mu)
\left(\ba{l} \hat{\mathcal{J}}^{\mu}_1 \\
             \hat{\mathcal{J}}^{\mu}_2 \\
             \hat{\mathcal{J}}^{\mu}_3 \ea\right) ,
\end{equation}
where $\vec{S}_\mu$ are the standard spin-1 matrices \cite{(Var88)},
which are generators of rotation in the vector representation. This shows
that the concept of alirotation (equivalent to changing the mixing of the
blocked state) translates into a change in the alignment  of
the system.

To illustrate how this works, let us examine the special case where the
states $|\mu\rangle\equiv|\mu_{y}\rangle$ and
$|\bar{\mu}\rangle\equiv|\bar{\mu}_{y}\rangle$ are eigenstates of the
$\hat{R}_{y}$ signature operator. Since:
\begin{equation}
\hat{R}_{i}\hat{R}_{j} = \sum_{k}\varepsilon_{ijk}\hat{R}_{k},
\end{equation}
we can express the states $|\mu_{y}\rangle$ and $|\bar{\mu}_{y}\rangle$
in terms of  $\hat{R}_{z}$-eigenstates:
\begin{equation}
\left( \ba{c}
|\mu_{y}\rangle \medskip\\
|\bar{\mu}_{y}\rangle
\ea\right)
=
\left( \ba{cc}
\displaystyle +\frac{1}{\sqrt{2}}e^{-i\pi/4} & \displaystyle +\frac{1}{\sqrt{2}}e^{-i\pi/4} \medskip\\
\displaystyle -\frac{1}{\sqrt{2}}e^{+i\pi/4} & \displaystyle +\frac{1}{\sqrt{2}}e^{+i\pi/4}
\ea\right)
\left( \ba{c}
|\mu_{z}\rangle \medskip\\
|\bar{\mu}_{z}\rangle
\ea\right),
\end{equation}
which corresponds to the rotation of the system by the Euler angles
$(\alpha,\beta,\gamma) = (0,\pi/2,\pi/2)$. Consequently, the alivector
$\mathcal{V}^{(\mu_{y})}$ is the vector $\mathcal{V}^{(\mu_{z})}$
alirotated by the angles $\gras{\phi}^{(\mu)} = (\alpha,\beta,\gamma)$.
Since (i) the blocked density matrix is not an aliscalar, and (ii)
alirotations are induced by rotations of the coordinate system or,
equivalently, a change of the symmetry operators used to label s.p.\
and q.p.\ states, we must conclude that the blocked density matrix may
depend on the choice of the symmetry operators that commute with the
Hamiltonian\footnote{Note that, in the particular case where the
alivector is built from the eigenstates $|\mu_{y}\rangle$ and
$|\bar{\mu}_{y}\rangle$ of $\hat{R}_{y}$, the alirotation by $(0,\pi,0)$
is equivalent to the $\hat{R}_{y}$ symmetry, and therefore leaves the
system invariant. This operation corresponds to
$(|\mu_{y}\rangle,|\bar{\mu}_{y}\rangle) \rightarrow
(+|\bar{\mu}_{y}\rangle,-|\mu_{y}\rangle)$. Therefore, in this particular
case, blocking state $|\mu_{y}\rangle$ or state $|\bar{\mu}_{y}\rangle$
gives exactly the same energy, even though time-reversal symmetry is
internally broken and the q.p.\ spectra do not exhibit the Kramers
degeneracy.}.

More generally, since all alignment properties of the system are
embedded in the vector-alivector operator $\hat{\mathcal{J}}$, we
also see that the alirotation of states $|\mu\rangle$ and
$|\bar{\mu}\rangle$  corresponds to {\HFB} states having different
alignment vectors. Therefore, the latter can be used to tag blocked
states. This is a very convenient method, which can be applied not
only for the time-even version of the functional when the quasiparticle
states $|\mu\rangle$ and $|\bar{\mu}\rangle$ are degenerate, but also
in the case of internally broken time-reversal symmetry. The key to
our considerations of  blocked states is the realization that blocking
must depend on the orientation of the alignment vector with respect
to the principal axes of the  mass distribution. Therefore, the only
rigorous way to proceed would be for each quasiparticle excitation
to vary the orientation of the alignment vector with respect to the
principal axes of the system, and retain the solution with the lowest
energy \cite{[Olb04]}. We give in Sec.~\ref{Sec-Results-Symmetries} a
pedagogical illustration of such anisotropy of blocking.

In many practical applications, however, one chooses a fixed direction
of alignment  dictated by practical considerations. In particular, the
identification of blocked single-particle states $n$ and quasiparticle
states $\mu_0$ is most conveniently carried out through the set of
conserved quantum numbers characteristic of the problem. In all calculations
performed in this work, nuclei are either axially deformed or nearly axial,
and they conserve reflection symmetry. The corresponding symmetry group is
$D_{2h}^{TD}$; hence, signature $r = \pm i$ and parity $\pi =\pm 1$ are
good quantum numbers. In {\HFODD} signature is defined with respect to
the $y$-axis of the reference frame \cite{[Dob04b]}. In this way, the alignment
vector is restricted to having only the $y$-component. To realize the three
possible alignments of the angular momentum along the principal axes, it is
sufficient to orient the longest, shortest, or intermediate axis along the
$y$-axis. Since in most cases, the configurations analyzed in this study
are axial, only two orientations suffice. We show in
Sec.~\ref{Sec-Method-Parameters} below how to implement such a scenario.

Equivalently, one could work with a good $z$-simplex basis such as in
Ref.~\cite{[Gir83]}. In that case, the default alignment is along the $z$-axis,
but results still depend on the orientation of the body. It is only by allowing
the alignment vector to cover the full solid angle that physical properties of
the system would not depend on the choice of the basis used to describe the odd
nucleus.


\section{Method of Calculation and Optimization Techniques}
\label{Sec-Method}

This section briefly describes the {\DFT} solvers used in this work
and discusses the choice of parameters entering our calculations. We
also outline various optimization techniques that have been implemented 
by us to carry out large-scale {\DFT} calculations for one-quasiparticle 
states on leadership class computers.


\subsection{Numerical Parameters}
\label{Sec-Method-Parameters}

All calculations in this work are performed with the {\DFT} solvers
{\HFBTHO} \cite{[Sto05]} and {\HFODD} \cite{[Dob04b],[Dob05],[Dob09d]}. Both
codes solve the Skyrme {\HFB} problem in the configuration space by
means of the Harmonic Oscillator ({\HO}) expansion technique. In {\HFBTHO},
the cylindrical {\HO} basis is used, and both axial and time-reversal
symmetries are imposed. This implies that the {\EFA} must be used for
blocking calculations. The 3D solver {\HFODD} employs the Cartesian {\HO}
basis and is symmetry-unrestricted. This unique feature of {\HFODD}
makes it a tool of choice for our study, since in the polarized
nuclear configurations many self-consistent symmetries are usually
broken. The blocking prescription is implemented exactly in {\HFODD}
with all the time-odd fields taken into account. Both codes have been
benchmarked against one another and they yield the same results
within a few eV for spherical or axially deformed even-even nuclei
\cite{[Dob04a]}.

As already mentioned, all nuclei considered in this work are either
axial or slightly triaxial, as well as reflection-symmetric. Therefore
$y$-signature and parity are conserved and used in {\HFODD} to tag q.p.\
and single-particle states. However, this implies that the total alignment
is confined to the $y$-axis. Since the latter is not the quantization
axis, one can not easily associate the single-particle spin $\Omega$
with the expectation value of the angular momentum: the situation is
analogous to the collective rotation in high-spin physics. For the sake
of identification of deformed Nilsson orbitals, it is convenient, however,
to reintroduce $\Omega$ as a (nearly) good quantum number by orienting
the angular momentum  along the $z$-axis; the resulting alignment properties
correspond to the limit of non-collective rotation.

To this end, we need to associate the quantization axis with the symmetry
axis of the nucleus. This can be achieved via a Euler rotation of the
body-fixed frame (by $\alpha = \pi/2$, $\beta = \pi/2$, $\gamma = 0$)
or by imposing constraints on the expectation values of the
quadrupole tensor $\hat{Q}_{2\mu}$. After testing these two options,
we choose the Euler rotation: calculations for even-even nuclei are
performed in the standard $y$-signature mode, then solutions are
Euler-rotated and used to warm-start calculations for odd-nuclei. In
this way, the Nilsson quantum number $\Omega$ is computed from the
expectation value of $\hat{j}_{y}$. This technique turns out to be
both stable and fast. Note that the energies of a given blocked state
in the Euler-rotated case and original orientation are different, as
discussed in Sec.~\ref{Sec-DFT-Symmetries}. Only a complete survey of
all possible orientations of the alignment vector, which would be a
major computational endeavour, could pin down the correct orientation.

As is well known, calculations for deformed nuclei converge faster if
the eigenstates are expanded on a stretched basis that follows the
geometry of the nuclear density. Unfortunately, the stretched basis
is not compatible with the Euler-rotation of the nucleus in space in
{\HFODD}. For that reason, all calculations presented in this work
have been carried out in a full spherical basis of $N_{\rm osc}$=14
oscillator shells (the number of basis states is $N_{s}$= 680). This
choice guarantees stability of results for the relatively modest
deformations considered in this study. The oscillator frequency was
fixed at $1.2\times\hbar\omega_{0}$ \cite{[Dob97]} for
$\hbar\omega_{0} = 41/A^{1/3}$\,MeV.

In this work, we use three commonly used Skyrme parametrizations:
SIII \cite{[Bei75]}, SkP \cite{[Dob84]}, and SLy4 \cite{[Cha98]}. In
the pairing channel, we employ the density-dependent delta
interaction in the mixed variant \cite{[Dob02c]}:
\begin{equation}
V(\gras{r}, \gras{r'}) =
V_{0}
\left[ 1 -
\frac{1}{2}\frac{\rho(\gras{r})}{\rho_0}
\right]\delta(\gras{r}-\gras{r'}),
\end{equation}
where $\rho_0=0.16$fm$^{-1}$ and  $V_{0}$ is the pairing strength
(identical for protons and neutrons). Note that the use of such a
zero-range interaction requires us to introduce a renormalization
(or regularization) procedure to avoid non-physical divergences
\cite{[Bul02],[Bor06fw]}. We employ the standard value
$E_{cut}$=60\,MeV.

For each Skyrme {\EDF}, the pairing strength $V_{0}$ has been adjusted
to reproduce the experimental proton odd-even mass difference in the
deformed nucleus $^{162}$Dy, $\Delta_{p}^{(3)} = 0.60$\,MeV. This
choice has been motivated by the findings of Ref.~\cite{[Ber09a]}
that by adjusting $V_{0}$ to experimental data for a spherical
semi-magic nucleus one underestimates pairing correlations in deformed
systems having lower single-particle level density around the Fermi
surface. Moreover, by considering the proton pairing gap, one effectively
takes into account the Coulomb contribution to pairing \cite{[Les08]}.
The pairing strengths used in this work are $V_{0} = -314.406$\,MeV,
$-297.303$ MeV, and $-249.059$\,MeV for SLy4, SIII, and SkP, respectively.
In Sec.\ref{Sec-Results-stability}, we also consider SkO,
$V_{0} = -269.226$\,MeV, and SkM*, $V_{0} = -297.875$\,MeV.


\subsection{Parallelization and Optimization}
\label{Sec-Method-Parallelization}

The advent of Teraflop and Petaflop supercomputers enables large-scale
surveys with symmetry-unrestricted {\DFT} solvers. To optimize resources,
however, optimization of the production codes is required. Starting from
the original published versions of {\HFODD} and {\HFBTHO}, we made a number
of improvements. First of all, a parallel interface using the standard
Message Passing Interface (MPI) has been constructed to allow the automated
distribution of calculations over several computing cores. Let us note that
the standard nuclear {\DFT} calculations are ``embarrassingly parallel".
Indeed, solving the {\HFB} equations for one nuclear configuration usually
does not exceed a few hours on a standard desktop computer. Therefore, each
computing core of a massively parallel system can process a single {\HFB}
task corresponding to a particular nucleonic configuration. Only in the limit
of very large {\HO} bases, or for {\DFT} solvers constructed in the coordinate
space, does the parallelization of the solvers become necessary. The
advantage of using massively parallel architectures is that simultaneous
calculations of hundreds or thousands of different many-body configurations
are possible in a very reasonable time. Such a strategy makes it possible to
extract systematic trends, use standard statistical analysis toolboxes, and
ultimately develop nuclear {\EDF}s of spectroscopic quality.

The scaling of a {\DFT} problem with the number of processors also implies
that a simple master-slave parallel architecture is sufficient for most
applications, and this solution is adopted here. All calculations in this
study were performed on the Cray XT3/XT4 Jaguar supercomputer at the National
Center for Computational Science at the Oak Ridge National Laboratory and on
the Cray XT3 Franklin supercomputer at the National Energy Research Scientific
Computing Center at the Lawrence Berkeley National Laboratory. Typical
production runs have involved from 8,000 to 12,000 computing cores per run,
and the typical calculation time was about 2 hours.

The {\HFB} equations represent a coupled system of non-linear equations for
nucleonic densities. The non-linearity enters through the dependence of the
mean fields on densities (self-consistency). In order to diminish the number
of iterations required to attain a given precision, we employ the modified
Broyden method \cite{[Joh88],[Bar08]}. The method is based on the observation
that the convergence of the {\HFB} process stops when the characteristic
variables in the problem, for example the density $\rho$, do not change any
more from one iteration to the next: $||\rho^{(n+1)} - \rho^{(n)}|| \approx 0$.
In other words, the {\HFB} equations can be viewed as a fixed-point problem,
and iterations can therefore be optimized by employing a quasi-Newton method.
It was shown that the computational cost (in units of number of iterations)
could be reduced substantially, by a factor of 3 to 4. Our particular
implementation of the modified Broyden method was described in Ref.~\cite{[Bar08]}.

As mentioned earlier, {\HFODD} solves the {\HFB} problem by expanding
eigenvectors in the Cartesian {\HO} basis. In terms of speed, one of the
major bottlenecks in {\HFODD} is the diagonalization of the {\HFB} matrix.
The latter is carried out with the subroutine \textsc{zheevr} of the {\LAPACK}
library. We found that a significant gain in terms of speed, up to 30-40\%
for large {\HO} bases with, e.g., $N_{\rm osc}$=20 shells, could be obtained
by using machine-specific implementations of the {\BLAS} and {\LAPACK}
libraries, such as {\ATLAS}.

\subsection{Massively Parallel Calculations: Convergence Improvements}
\label{Sec-Method-Convergence}

Based on the experience gained in self-consistent calculations for odd-$A$ 
nuclei, it appears that calculations involving blocking are always less 
stable than those performed for even-even nuclei. Apart from the specific 
issue related to finite-size instabilities addressed in 
Sec.~\ref{Sec-Results-stability} below, these numerical instabilities
are related to the need to select, at each iteration, the blocked q.p.\ 
state. The blocking procedure is outlined in Sec.~\ref{Sec-DFT-Blocking} 
and the detailed justification can be found in, e.g., Ref.~\cite{[Hee95]}. 
The selection method implies that the blocked q.p.\ state may change from 
one iteration to the next, in particular at the beginning of the 
calculation. This numerical noise is the price to pay for the full 
self-consistency, and it explains why small differences in the initial 
conditions can actually affect the convergence process.

When only a few nuclei are considered, and a small number of blocked
configurations near the ground-state is calculated, one can often
find ways to converge calculations, such as:
(i) changing the linear-mixing parameter of the self-consistent scheme;
(ii) starting from the unblocked fully-paired state corresponding to an 
odd average particle number (false vacuum) \cite{[Dug01a]};
(iii) starting from the even-even nucleus with one more particle for a
particle-like blocked state and with one particle less for a
hole-like blocked state, as implemented in Ref.~\cite{[Hee95]}
(we also used this method in our calculation);
(iv) using different values of the linear-mixing parameter for time-even,
time-odd, and/or pairing fields.
Whenever a blocking calculation fails to converge, one may repeat it by 
using one or several of these tricks until a converged result is obtained. 
This is what was done in previous studies involving self-consistent 
calculations, and it was possible, because these studies were focused
on ground-state properties and only a minimum number of different
configurations was considered.

\begin{table*}[ht]
\caption{Comparison of  {\EFA} ({\HFBTHO}) with exact blocking ({\HFODD})
for four one-quasineutron configurations in $^{121}$Sn. The time-odd fields
are switched off. Shown are: the quasiparticle energy $E_{qp}$, neutron
chemical potential $\lambda_{n}$, neutron pairing energy $E_{pair}^{n}$,
average neutron pairing gap $\Delta_{n} = \text{Tr}(\Delta\rho)/N$, total
r.m.s. radius, axial quadrupole deformation $\beta$, total quadrupole moment
$Q_{tot}$, kinetic energy $E_{kin}$ (for protons and neutrons), total
spin-orbit energy $E_{SO}$, direct Coulomb energy $E_{dir}$, and total energy
$E_{tot}$. The last two lines show the {\HFODD} alignments of the blocked
quasiparticles: $J_{\parallel}$ was calculated in the non-collective orientation
and $J_{\perp}$ in the collective orientation, see Discussion in Sec.
\ref{Sec-Results-Symmetries}. In the EFA total alignments are equal to zero
by construction. The orbits are labelled by the $\ell_j(\Omega^\pi)$ quantum
numbers. The SLy4 functional is used in the particle-hole channel and the
density-dependent delta interaction with  $V_{0}$=--285.634\,MeV. The differences
between {\HFBTHO} and {\HFODD} results are highlighted in boldface.}
\begin{tabular}{cdddddddd}
\hline\hline                          \text{Exact}
        State        & \multicolumn{2}{c}{$d_{3/2}(1/2^{+})$} & \multicolumn{2}{c}{$d_{3/2}(3/2^{+})$} & \multicolumn{2}{c}{$h_{11/2}(5/2^{-}$)} & \multicolumn{2}{c}{$g_{7/2}(7/2^{+})$}\\
        Blocking     &  \multicolumn{1}{c}{\EFA} & \multicolumn{1}{c}{\text{Exact}} & \multicolumn{1}{c}{\EFA} & \multicolumn{1}{c}{\text{Exact}} & \multicolumn{1}{c}{\EFA} & \multicolumn{1}{c}{\text{Exact}} & \multicolumn{1}{c}{\EFA} & \multicolumn{1}{c}{\text{Exact}}  \\
\hline
   $E_{qp}$  (MeV)   &    1.00{\bf 76} &    1.0080 &     1.182{\bf 2} &    1.1820 &     1.4570       &    1.4570 &    2.28{\bf 79} &    2.2880 \\
 $\lambda_{n}$ (MeV) &   -7.7496       &   -7.7496 &    -7.7288       &   -7.7288 &    -7.983{\bf 4} &   -7.9836 &   -7.6371       &   -7.6371 \\
$E_{pair}^{n}$ (MeV) &   -9.294{\bf 9} &   -9.2948 &    -9.441{\bf 1} &   -9.4410 &    -8.714{\bf 5} &   -8.7147 &  -10.40{\bf 41} &  -10.4036 \\
 $\Delta_{n} $ (MeV) &    1.0575       &    1.0575 &     1.0667       &    1.0667 &     1.0395       &    1.0395 &    1.1206       &    1.1206 \\
   r.m.s (fm)        &    4.6895       &    4.6895 &     4.6889       &    4.6889 &     4.6894       &    4.6894 &    4.6884       &    4.6884 \\
   $\beta$           &   -0.0257       &   -0.0257 &     0.0131       &    0.0131 &     0.009{\bf 9} &    0.0098 &    0.0340       &    0.0340 \\
  $Q_{tot}$ (b)      &   -0.862{\bf 7} &   -0.8624 &     0.4383       &    0.4383 &     0.33{\bf 01} &    0.3351 &    1.1409       &    1.1409 \\
$E_{kin}^{n}$  (MeV) & 1360.438{\bf 5} & 1360.4384 &  1360.9970       & 1360.9970 &  1358.89{\bf 95} & 1358.8980 & 1364.56{\bf 96} & 1364.5680 \\
$E_{kin}^{p}$  (MeV) &  827.317{\bf 6} &  827.3177 &   827.3582       &  827.3582 &   827.189{\bf 2} &  827.1890 &  828.1830       &  828.1830 \\
$E_{SO}^{tot}$ (MeV) &  -50.4839       &  -50.4839 &   -50.8174       &  -50.8174 &   -49.58{\bf 56} &  -49.5844 &  -54.65{\bf 31} &  -54.6529 \\
$E_{dir}$      (MeV) &  365.7437       &  365.7437 &   365.7638       &  365.7638 &   365.738{\bf 7} &  365.7386 &  365.99{\bf 80} &  365.9979 \\
$E_{tot}$      (MeV) &-1024.707{\bf 4} &-1024.7073 & -1024.6538       &-1024.6538 & -1024.385{\bf 6} &-1024.3855 &-1023.4465       &-1023.4465 \\
$J_\parallel$ ($\hbar$)&  0.00         &    0.50   &     0.00         &    1.50   &     0.00         &    2.50   &    0.00         &    3.50   \\
$J_\perp$     ($\hbar$)&  0.00         &    0.82   &     0.00         &    0.00   &     0.00         &    0.43   &    0.00         &    0.01   \\
\hline\hline
\end{tabular}
\label{table01}
\end{table*}
In our case, however, we consider thousands of configurations and
such a trial-and-error scheme, however helpful, is simply impossible
to implement. Instead, we have resorted to a simple trick; namely, since 
the initial conditions do matter, we artificially generate slightly 
different initial conditions for the two signature partners that 
converge to practically the same result. 

The main idea here consists in adding a tiny rotational frequency of
about $\hbar\omega$=0.001\,MeV to break the degeneracy of signature
configurations. This improves  the convergence rate at the price of 
an insignificant numerical error of about 1--2\,keV, on average. We 
illustrate this fact by two specific examples of blocked states in 
$^{163}$Tb (native version, non-collective orientation) with different 
alignments. For the blocked state [411]1/2, at $\hbar\omega$=0 the 
total energy equals $-$1322.279268\,MeV, while for $\hbar\omega$=+0.001 
and $-$0.001\,MeV the total energies read $-$1322.279480\,MeV 
($J_{\parallel} = -\Omega = -1/2$) and $-$1322.279188\,MeV 
($J_{\parallel} = +\Omega = +1/2$), respectively. Similarly, for the 
blocked state [404]7/2, the three corresponding energies are: $
-$1321.725322\,MeV, $-$1321.725538\,MeV ($J_{\parallel} = -\Omega = 
-7/2$), and $-$1321.726010\,MeV ($J_{\parallel} = +\Omega = +7/2$). 
To make our point, we deliberately show these energies with far 
too many digits than it is physically relevant. In these particular two 
examples, the numerical precision of the calculation is 10\,eV, which is 
exceptionally good for odd nuclei. Therefore, the noted differences can 
only be attributed to the effect of the cranking term. Without this term, 
by blocking states $\mu_{0}$ and $\bar{\mu}_{0}$ one always obtains exactly
the same {\HFB} energy, $E_{\mu_{0}} = E_{\bar{\mu}_{0}}$, because the full
Skyrme functional is time-even.


\section{Results}
\label{Sec-Results}

This section presents a number of Skyrme-{\HFB} results for odd-mass nuclei.
We begin by giving a detailed numerical comparison of the {\EFA} with the exact
blocking prescription in the limit of conserved time-reversal symmetry. The
impact of time-odd fields on the quasiparticle spectrum in the rare-earth region
is shown in Sec.~\ref{Sec-Results-Todd} with the native, gauge, and Landau
versions of SIII, SkP, and SLy4 functionals. The role of the nuclear alignment
vector on physical observables is also studied. Results of calculations are
compared with selected experimental data in Sec. \ref{Sec-Results-Experiment}.
The triaxial polarization induced by a blocked quasiparticle is discussed in
Sec.~\ref{Sec-Results-Gamma}. Finally, Sec. \ref{Sec-Results-stability} mentions
the problem of the intrinsic instability of certain Skyrme functionals that appear
when time-odd terms are included.


\subsection{Validation of the EFA Approximation}
\label{Sec-Results-EFA}

To demonstrate the numerical precision of our calculations, Table
\ref{table01} shows the results for four one-quasineutron states in
$^{121}$Sn obtained with {\HFBTHO} ({\EFA}) and {\HFODD} (exact
blocking). They are selected based on the mean-field spectrum
(\ref{equivalent-spectrum}) of $^{120}$Sn. For the sake of this
comparison, the time-odd fields in {\HFODD} have been switched off,
thereby enforcing the regime where the exact blocking procedure is
strictly equivalent to the {\EFA}, see Sec. \ref{Sec-DFT-Blocking}.
Indeed, the obtained numerical differences between the {\EFA} and
exact blocking are extremely small, less than 1\,keV for the four
different cases shown in Table \ref{table01}. This can be entirely
attributed to various implementations adopted differently in the two
codes such as the method of computing the Coulomb potential, etc. The
even-even core $^{120}$Sn is spherical in its ground state. The
quasiparticle blocking slightly polarizes the nuclear shape inducing
small quadrupole deformations for some configurations.

Although the time-even observables obtained within the {\EFA} and
exact-blocking  are strictly identical if the time-odd fields are
disregarded, this is not true for time-odd observables. In Table
\ref{table01}, this is illustrated by the values of alignments of
the blocked quasiparticles aligned parallel ($J_\parallel$) or
perpendicular ($J_\perp$) to the symmetry axis. Of course, without
time-odd fields, the direction of alignment does not influence the
time-even observables.

\begin{table}[ht]
\begin{center}
\caption{Comparison of {\EFA} ({\HFBTHO}) with exact blocking
({\HFODD}) for ten one-quasiproton configurations in $^{163}$Tb.
Shown are the total energy $E_{tot}$ and total quadrupole moment
$Q_{tot}$. The orbits are labeled by the asymptotic Nilsson quantum
numbers $[Nn_z\Lambda]\Omega^\pi$. The SIII Skyrme functional is
used in the particle-hole channel. Time-odd fields are disregarded.
The differences between {\HFBTHO} and {\HFODD} results are shown in
boldface.
}
\begin{ruledtabular}
\begin{tabular}{ccccc}
               & \multicolumn{2}{c}{{\EFA} ({\HFBTHO})} & \multicolumn{2}{c}{Exact ({\HFODD})} \\
Blocked State  & $Q_{tot}$ (b) & $E_{tot}$ (MeV)  & $Q_{tot}$ (b) & $E_{tot}$ (MeV) \\
\hline
 $[411]3/2^+$ &  18.514       & $-$1323.495       & 18.514       & $-$1323.495 \\
 $[532]5/2^-$ &  17.759       & $-$1322.64{\bf 8} & 17.759       & $-$1322.64{\bf 7} \\
 $[523]7/2^-$ &  18.55{\bf 4} & $-$1322.41{\bf 5} & 18.55{\bf 5} & $-$1322.41{\bf 4} \\
 $[411]1/2^+$ &  18.384       & $-$1322.322       & 18.384       & $-$1322.322 \\
 $[413]5/2^+$ &  18.654       & $-$1322.151       & 18.654       & $-$1322.151 \\
 $[541]1/2^-$ &  20.13{\bf 8} & $-$1321.77{\bf 1} & 20.13{\bf 6} & $-$1321.77{\bf 3} \\
 $[541]3/2^-$ &  17.29{\bf 1} & $-$1321.357       & 17.29{\bf 0} & $-$1321.357 \\
 $[530]1/2^-$ &  17.03{\bf 4} & $-$1320.762       & 17.03{\bf 2} & $-$1320.762 \\
 $[420]1/2^+$ &  17.76{\bf 6} & $-$1320.090       & 17.76{\bf 7} & $-$1320.090 \\
 $[404]9/2^+$ &  19.266       & $-$1319.851       & 19.266       & $-$1319.851 \\
\end{tabular}
\label{table02}
\end{ruledtabular}
\end{center}
\end{table}

Rare-earth nuclei provide an excellent testing ground for studies of
deformed Nilsson orbits. Many of those nuclei are well-deformed,
near-axial rotors and the deformed mean field theory is particularly
suitable to describe their structural properties. Table \ref{table02}
shows a comparison for several one-quasiproton configurations in a
well deformed odd-proton nucleus $^{163}$Tb. In {\HFBTHO}, the
determination of a  blocking candidate was made using the mean-field
spectrum (\ref{equivalent-spectrum}) of the even-even core $^{162}$Dy.
In the case of {\HFODD}, to improve speed and stability of the
iterative process \cite{[Hee95]}, blocking candidates of a particle
character (above the proton Fermi level of $^{162}$Dy) were selected
from the mean-field spectrum of $^{164}$Dy while hole-like levels were
selected from that of $^{162}$Dy. Of course, the final results do not
depend on which particular even-even nucleus has been  used as a core.

The results shown in Table~\ref{table02} show again that without
time-odd fields, the full blocking procedure is equivalent to the
{\EFA}. It is worth noting that the difference on the total energy
is less than 1\,keV for all of the excited states, and less than
0.002\,b for the quadrupole moments, regardless of the quadrupole
polarization exerted by a blocked state.


\subsection{Effect of Time-Odd Fields}
\label{Sec-Results-Todd}

This section illustrates the effect of the various prescriptions
for the time-odd channel (\ref{Hodd}) of the functionals. Calculations
were performed for all odd-proton nuclei with 63$\le$$Z$$\le$75 and
78$\le$$N$$\le$104. For each of them, 14 non-degenerate blocked
configurations around the Fermi level have been considered.
Altogether, 3,822 independent one-quasiproton states were studied.

\subsubsection{Native Functionals}

Table~\ref{table03} displays results for one-quasiproton states
in $^{163}$Tb in the time-even, native, gauge, and Landau variants
of calculations. The alignment and elongation axes coincide with
the $y$-axis of the reference frame. The time-even energies are
shown in the absolute scale. For other variants, shown are
displacements with respect to the time-even case:
\begin{equation}
\Delta E^{\text{Todd}} = E_{qp}^{\text{Todd}=0} - E_{qp}^{\text{Todd}\neq0}.
\label{Etodd}
\end{equation}
In the particular example shown in Table~\ref{table03}, the maximum
shift in one-quasiparticle levels due to the time-odd terms of the
native functional is 127\,keV. This is consistent with the earlier
results of Refs.~\cite{[Dug01a],[Dug01b]} and overall smaller than
in the relativistic mean field approach, where time-odd polarization
corrections seem to range from only a few dozen of keV in deformed
actinides up to half a MeV in light nuclei \cite{[Afa03],[Afa10]}.
\begin{table}[ht]
\begin{center}
\caption{Energies (in MeV) of one-quasiproton configurations in $^{163}$Tb
calculated using the time-even, native, gauge, and Landau variants of the
SIII Skyrme functional, see Sec.~\ref{Sec-DFT-EDF}. Results for the time-even
variant are shown in the absolute scale, while those for the other variants
are shown as shifts (\ref{Etodd}).
}
\begin{ruledtabular}
\begin{tabular}{ccccc}
Blocked State & Time-even & Native  &  Gauge  & Landau \\
\hline
$[411]3/2^+ $ & $-$1323.495 & $-$0.075 & +0.042 & $-$0.125 \\
$[532]5/2^- $ & $-$1322.647 & $-$0.052 & +0.029 & $-$0.105 \\
$[523]7/2^- $ & $-$1322.410 & $-$0.060 & +0.039 & $-$0.080 \\
$[411]1/2^+ $ & $-$1322.322 & $-$0.043 & +0.040 & $-$0.118 \\
$[413]5/2^+ $ & $-$1322.151 & $-$0.048 & +0.062 & $-$0.085 \\
$[541]1/2^- $ & $-$1321.773 & $-$0.007 & +0.055 & $-$0.075 \\
$[541]3/2^- $ & $-$1321.357 & $-$0.047 & +0.036 & $-$0.107 \\
$[530]1/2^- $ & $-$1320.762 & $-$0.037 & +0.017 & $-$0.161 \\
$[420]1/2^+ $ & $-$1320.090 & $-$0.127 & +0.018 & $-$0.231 \\
$[404]9/2^+ $ & $-$1319.851 & $-$0.121 & +0.036 & $-$0.150    \\
\end{tabular}
\label{table03}
\end{ruledtabular}
\end{center}
\end{table}

\begin{figure}[ht]
\includegraphics[width=1.0\columnwidth]{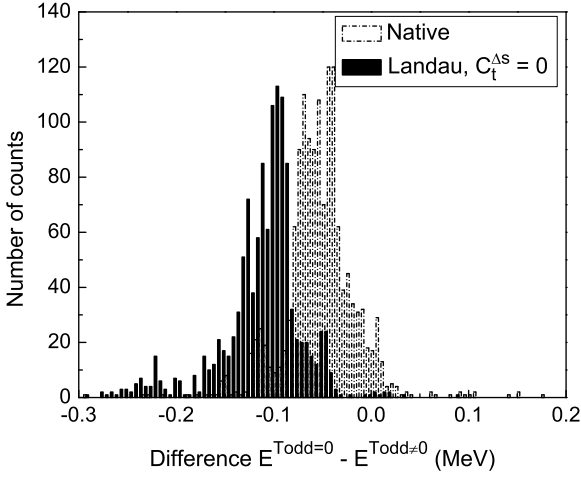}
\caption{Cumulative histogram of energy differences (\ref{Etodd}) for
one-quasiproton states in the deformed rare-earth nuclei calculated with
SIII, SkP, and SLy4 {\EDF}s. Dot-dashed open bins: native functionals;
solid filled bins: Landau functionals. The bin size is 5\,keV.
}
\label{fig01}
\end{figure}

\begin{figure}[ht]
\includegraphics[width=0.95\columnwidth]{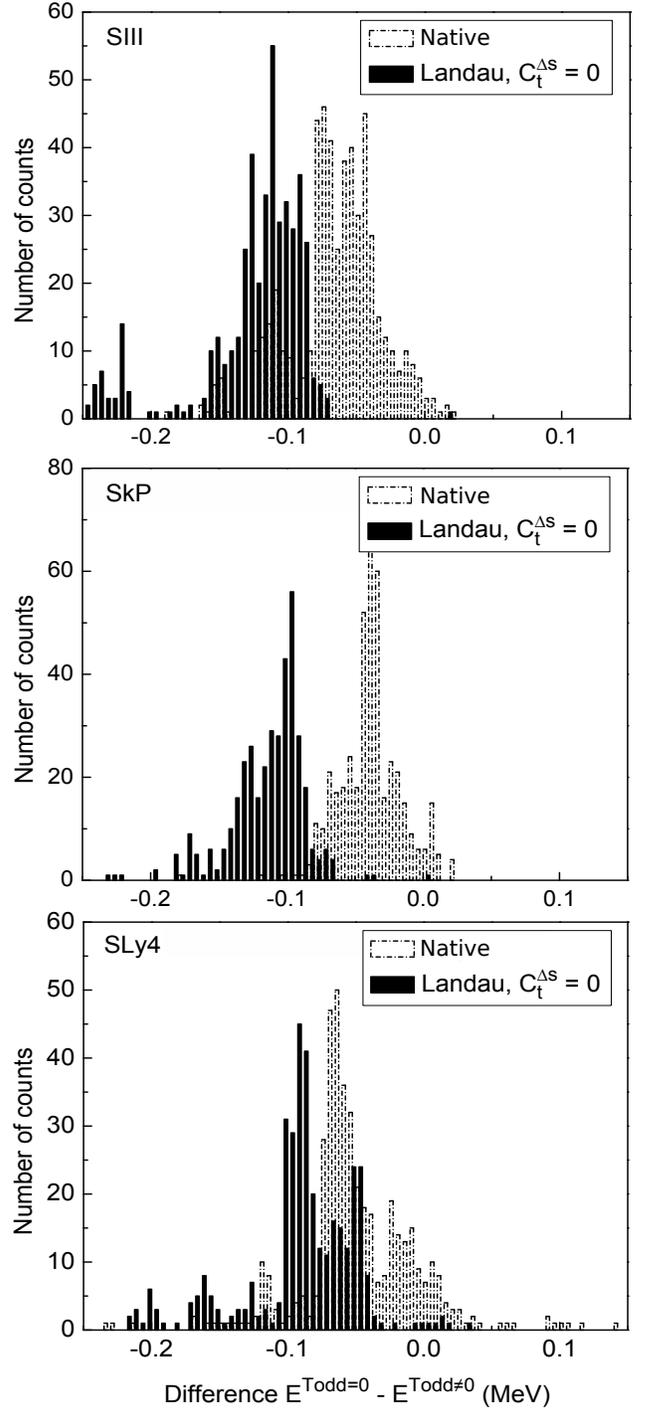}
\caption{Similar to Fig.~\ref{fig01} except for individual Skyrme
functionals: SIII (top), SkP (middle), and SLy4 (bottom).
}
\label{fig02}
\end{figure}

The overall impact of the time-odd fields on the energy of
one-quasiproton states in the deformed rare-earth nuclei is
summarized in Fig.~\ref{fig01} which  shows the distribution of
$\Delta E^{\text{Todd}}$ (\ref{Etodd}) for SIII, SkP, and SLy4 {\EDF}s.
When native functionals are used, the total number of converged cases
is 1,404 (524 for  SIII, 443 for SkP, and 437 for SLy4). The average
value of $\Delta E^{\text{Todd}}$ is --50\,keV with a standard
deviation of 42\,keV.

The magnitude of the odd-time effect depends on the choice of the
{\EDF}. To illustrate this point, Fig.~\ref{fig02} displays the
distribution of $\Delta E^{\text{Todd}}$ for individual functionals.
Focusing in this section on the native functionals (dot-dashed open
bins), it is seen that the largest time-odd effect is predicted for
SIII, which also shows an appreciable spread in values (configuration
dependent). On the other hand, for the SkP parametrization the
distribution of $\Delta E^{\text{Todd}}$  is fairly narrow, centered
around --40\,keV.

\begin{figure}[ht]
\includegraphics[width=1.0\columnwidth]{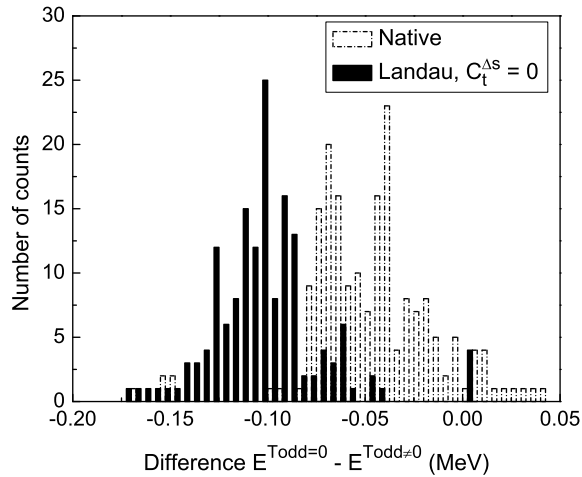}
\caption{Similar to Fig.~\ref{fig01} except for g.s configurations
only.
}
\label{fig03}
\end{figure}

By construction, Figs.~\ref{fig01} and \ref{fig02} contain
contributions from g.s.\ configurations and from nearly-lying
excited states. Since g.s values are of particular importance as they
impact mass predictions, Fig.~\ref{fig03} shows $\Delta
E^{\text{Todd}}$ for g.s.\ configurations only. The average value of
the g.s.\ time-odd displacement is only $\sim$50\,keV. Most of the
few cases with $|\Delta E^{\text{Todd}}_{\text{gs}}| > 150$\,keV
correspond in fact to a collapse of pairing correlations in one of
the 2 sets of calculations. It may be worth noting that the most
recent {\HFB} mass formula based on the Skyrme BSk17 parametrization
yields a r.m.s deviation of 581\,keV \cite{[Gor09]}. The uncertainty
associated with neglecting the time-odd fields appears, therefore,
to be smaller by an order of magnitude.

\begin{figure}[ht]
\includegraphics[width=0.9\columnwidth]{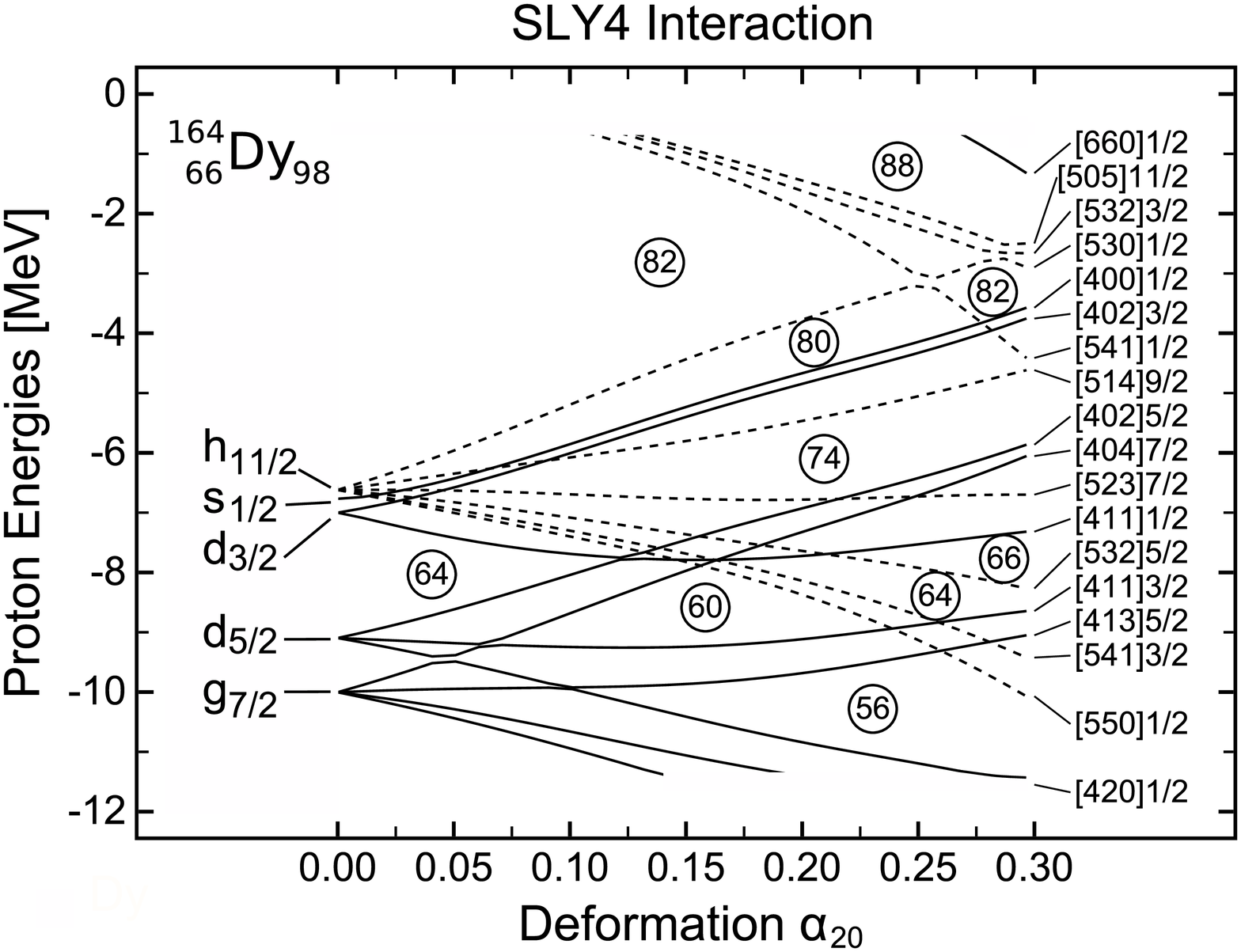}
\caption{Nilsson proton levels of SLy4 in $^{164}$Dy as function of
the axial quadrupole deformation $\alpha_{20}$ ($\alpha_{22}=0$).
}
\label{fig04}
\end{figure}

In order to discuss the configuration dependence of the time-odd
displacement, it is instructive to identify the s.p.\ orbits of
interest. To this end, Fig.~\ref{fig04} shows the evolution of the
proton s.p.\ energies, defined as the eigenvalues of the mean field
operator (\ref{equivalent-spectrum}), in the nucleus  $^{164}$Dy
calculated with SLy4 as a function of the axial quadrupole deformation
$\alpha_{20}$. This Nilsson diagram has been obtained by carrying out
a set of constrained {\HFB} calculations along a one-dimensional
$\langle \hat{Q}_{20}\rangle$ path.

Although the values of $\Delta E^{\text{Todd}}$ are usually small,
there are a few cases where the displacement can amount to more than
100\,keV. A detailed analysis of the blocked configurations for all
three interactions shows that the largest deviations correspond
essentially to the [420]1/2, [404]9/2, [400]1/2, and [505]11/2 Nilsson
orbitals. It is interesting to note that the value of the s.p.\ angular
momentum does not seem to be crucial, since these states can be
associated with both low and high-$j$ spherical shells. In deformed
rare-earth nuclei, equilibrium deformations are
$\alpha_{20} \sim 0.25-0.30$. As seen in Fig.~\ref{fig04}, in this
deformation range, the orbital [420]1/2 is a deep-hole state while
[404]9/2, [400]1/2, and [505]11/2 are highly excited particle states.
All these one-quasiproton excitations are strongly oblate-driving. A
similar result has also been obtained for SLy4 and SkP.

\subsubsection{Landau functionals}

Traditionally, only the time-even channel of Skyrme functionals has been
adjusted to selected experimental data. That is, the time-odd channel
has usually not been constrained.  This is illustrated by the broad spread
of the values of the isoscalar Landau parameters $g_{0}$ and $g_{1}$
of the standard Skyrme functionals \cite{[Ben02],[Zdu05]}. In
\cite{[Ben02]}, a careful study of Gamow-Teller resonances within the
Skyrme EDF theory yielded a set of `optimal' Landau parameters that
could be used to fix some of the coupling constants of the time-odd
channel of the functional (namely the $C_{t}^{s}$ and $C_{t}^{T}$).

As seen in Table~\ref{table03}, the time-odd polarization in the Landau
variant is greater than in the native variant, with the largest shift
growing to  231\,keV. The time-odd shifts in the gauge variant are generally
smaller than for the native and Landau parameterizations. They also have
opposite sign (time-odd polarization in the gauge variant decreases the
binding energy while it is repulsive in native and Landau variants).

The solid-filled bins in Figs.~\ref{fig01}--\ref{fig03} show
$\Delta E^{\text{Todd}}$ for Landau-corrected functionals. The effect
of this correction is significant, as it shifts the centroid of most
histograms by about 100\,keV for SIII and SkP and 50\,keV for SLy4.
When only ground-states are considered, the overall shift is of the
order of 50\,keV.

To finish this section, let us recall that setting $C_{1}^{\Delta s}=0$
was motivated in \cite{[Ben02]} to reproduce the energy and strength of
the GT resonance, although different conclusions about the role of this
term were obtained later in \cite{[Fra07]}. In any case, the isoscalar
channel governed by the $C_{0}^{\Delta s}$ term is not constrained by
GT resonances, and in Refs.~\cite{[Ben02],[Zdu05],[Sat08],[Zal08]}
these terms are set to zero essentially to ensure the stability of
the calculation. We briefly discuss this point in
Sec.~\ref{Sec-Results-stability}.

\subsubsection{Alignments and Choice of the Quantization Axis}
\label{Sec-Results-Symmetries}

As discussed in Sec.~\ref{Sec-DFT-Symmetries}, one of the
characteristic features of the treatment of odd nuclei in the
blocking approximation is the dependence of time-odd densities on the
orientation of the alignment vector with respect to the principal
axes or, equivalently, the choice of the self-consistent symmetries
and quantization axis. To measure this effect, we performed two sets
of calculations. The first variant ($\perp$) corresponds to the
alignment vector aligned along the $y$-axis, and the shape symmetry
axis aligned along the $z$-axis. In the terminology of the cranking
model, this case represents ``collective rotation'' perpendicular
to the symmetry axis. In the second variant ($\parallel$), the
nucleus is rotated by 90$^{\text{o}}$, as described in
Sec.~\ref{Sec-Method-Parameters}, so that the alignment and
symmetry axes coincide with the $y$-axis (``non-collective rotation'').

\begin{figure}[ht]
\includegraphics[width=\columnwidth]{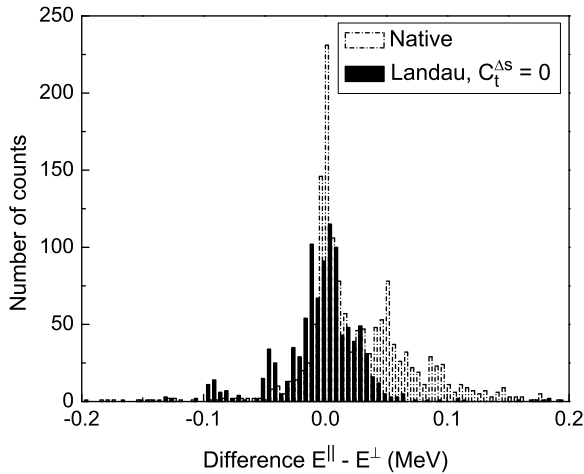}
\caption{Cumulative histogram of deviations
$\Delta E^j = E_{qp}^{\parallel} - E_{qp}^{\perp}$ between energies
of one-quasiproton states calculated in the non-collective-rotation
($\parallel$) and collective-rotation ($\perp$) variants. Dot-dashed
open and solid filled bins correspond, respectively, to native
and Landau variants. The bin size is 5\,keV. }
\label{fig05} \end{figure}

Note that in both situations $y$-signature and parity are conserved:
the identification of blocking configurations via the position of the
blocked state in a given signature/parity block hence provides a
very robust way of tracking configurations before and after the
Euler-rotation, as it is independent of the changes in other {\it spatial}
characteristics of the quasi-particle wave-functions. As mentioned in
Sec.~\ref{Sec-Method-Parameters}, the original Nilsson label of a q.p.
state can be easily recovered in the non-collective orientation by simply
exchanging the roles of the $z$ and $y$ axis in their computation.

In Fig.~\ref{fig05} we show the distribution of differences
$\Delta E^j = E_{qp}^{\parallel} - E_{qp}^{\perp}$ for the 3,822
cases presented in the previous section (only those well converged
are included in the plot). It is seen that the time-odd polarization
due to the orientation of qp alignment gives an appreciable contribution
to the time-odd shift, with the average value of $\Delta E^j$ being
about 50\,keV in the native variant. The orientation effect seems to
be weaker for Landau functionals. While the energy shift $\Delta E^j$
depends on the actual configuration, the total energy in the collective
rotation scenario ($\perp$) is overall lower than in the non-collective
one ($\parallel$) when native functionals are used.


\subsection{Experimental Odd-proton Spectra}
\label{Sec-Results-Experiment}

In well-deformed  nuclei, one quasiparticle states can be related
to the rotational band-head configurations \cite{(Boh75)}. In
rare-earth nuclei, rich systematics of experimental data exist,
and most importantly, the customary assignments of Nilsson labels
$[Nn_{z}\Lambda]\Omega$ are available \cite{[Naz90],(ensdf)}.
Although these labels are approximate, they facilitate the
comparison between theory and experiment.

\begin{figure*}[ht]
\includegraphics[height=0.75\textwidth]{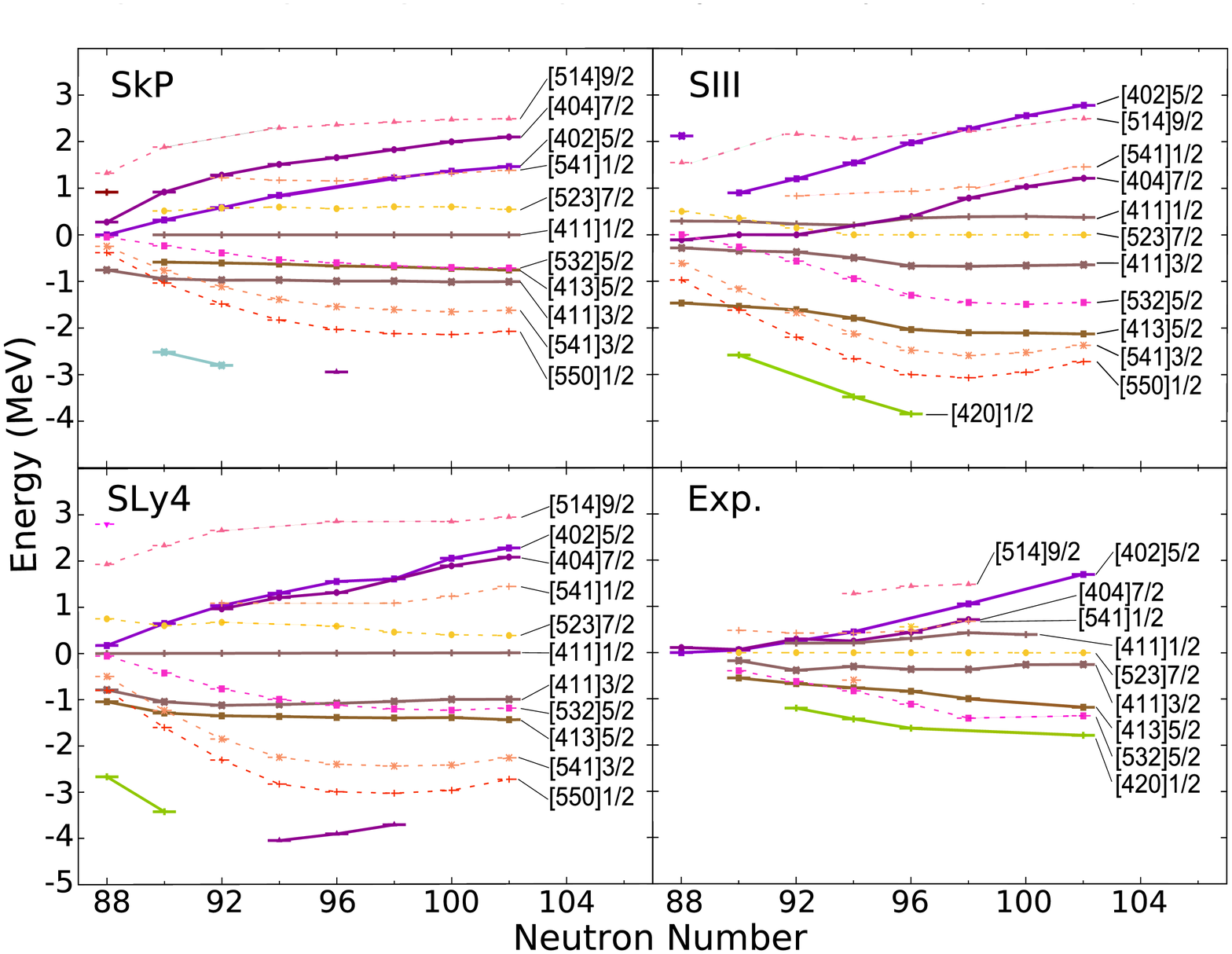}
\caption{One-quasiproton band-head energies: calculated with SkP
(top left), SIII (top right), and SLy4 (bottom left) Skyrme
functionals, and extracted from experimental data (bottom right), plotted
versus neutron number for Ho isotopes. Hole states are plotted below
the ground state (zero energy) and particle states are plotted above.
The time-odd terms are included in the native variant. Experimental
data are taken from Refs.~\cite{[Naz90],(ensdf)}.
}
\label{fig07}
\end{figure*}

In Fig.~\ref{fig07} we show the one-quasiproton spectra
for the Ho ($Z$=67) isotopic chain predicted with  SkP (upper-left
panel), SIII (lower-left panel), and SLy4 (lower-right panel)
functionals in the native variant. They are compared to experimental data.
We follow the convention of Refs.~\cite{[Nie75],[Naz90]}, whereby the
hole-like excitations are plotted below zero (representing the g.s.\
configuration) while the particle-like states are plotted above zero.

The comparison with experiment  suggests that the functional
parametrizations employed in our work are not of spectroscopic
quality for deformed nuclei. While the general deformation trends
are reproduced and most of the orbitals found experimentally are
indeed predicted to appear  around the Fermi level, the
quantitative agreement with the data is not particularly impressive.
For example, the SLy4 parametrization  fails to reproduce the observed
[523]7/2 g.s.\ of Ho isotopes; this state is predicted to lie
300--500\,keV above the calculated [411]1/2 ground state.
Surprisingly, the oldest Skyrme parametrization SIII gives the
best reproduction of experimental band heads. The result
of Fig.~\ref{fig07} is consistent with the conclusions of
Ref.~\cite{[Bon07]}; they found that the agreement of both spin and
parity in the self-consistent models reaches about 40\% for
well-deformed nuclei regardless of the Skyrme force used.

The three functionals used here have different isoscalar effective
masses, $m^{*}$=1, 0.707, and 0.7 for SkP, SIII, and SLy4,
respectively. The effect of $m^*$ on shell structure is complex
\cite{[Kor08]}; among others, it impacts the density of states
around the Fermi level. As seen in Fig.~\ref{fig07}, the average
level density obtained with SkP is indeed close to the experimental
one. However, this does not necessarily mean that the spectroscopic
properties are better described with this interaction: just as for
SLy4, the ground-state is incorrectly assigned to the [411]1/2
orbital for all isotopes.

There are, indeed, many factors that may impact the order of
one-quasiparticle states. The recent analysis of spherical s.p.\ shell
structure \cite{[Kor08]} has demonstrated that the isoscalar coupling
constants in {\EDF} have a large impact on the position of s.p.\ energies
and spin-orbit splitting. It was also shown that the role of the
effective-mass coupling constant cannot be reduced to merely changing
the overall density of states. In fact, effective mass significantly
influences relative positions of single-particle levels, including the
splitting of spin-orbit partners.


\subsection{Triaxial Shape Polarization}
\label{Sec-Results-Gamma}

Triaxial deformations of nuclear shape are enhanced  at high spins
\cite{(Szy83),[Fra00]}. One spectacular example is the nuclear wobbling
motion, which is caused by the fast rotation of triaxially-deformed nuclei
\cite{(Boh75),[Sho09],[Ode01w],[Jen02w]}. The phenomena of nuclear chirality
is also tightly related to axial asymmetry \cite{[Fra97],[Olb04],[QiZ09]}.
Recently, a systematic study of ground-state nuclear shapes in the framework
of the macroscopic-microscopic model has also pointed to regions of
triaxial instability in the nuclear chart \cite{[Mol06]}.

In the deformed rare-earth region that we consider in this work, the
blocking of a quasiparticle built on intruder configurations has a strong
$\gamma$-driving effect \cite{[Fra83],[Ham83w],[Abe90],[Mat07]}. Most of
the studies of this phenomenon are so far confined to high-spin states.
Our calculations offer the opportunity to assess the degree of triaxiality
in the g.s. configurations associated with weakly spin-polarized states.

\begin{figure}[ht]
\includegraphics[height=0.95\columnwidth ,angle=-90]{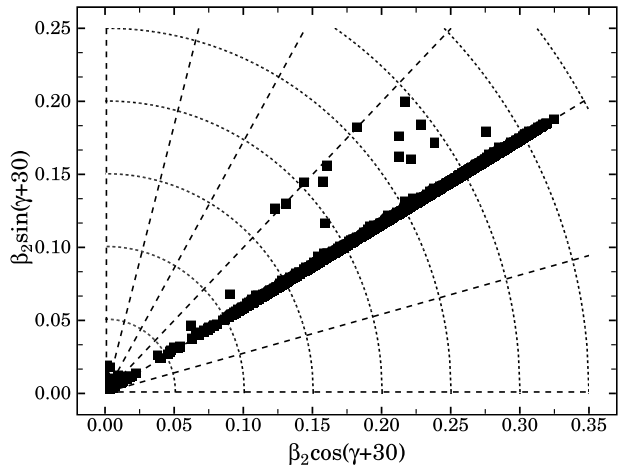}
\caption{Distribution of the equilibrium deformations in the
$(X,Y)$ plane, $X = \beta\cos(\gamma + \pi/6)$ and
$Y = \beta\sin(\gamma+\pi/6)$, where $\beta$ and $\gamma$ are
the standard Bohr quadrupole deformation parameters. The figure
corresponds to the 3,822 different blocked configurations
considered in this study. Time-odd terms are set to zero.
  }
\label{fig08}
\end{figure}

The calculated equilibrium deformations of one quasi-proton
configurations considered in our survey are displayed in Fig.~\ref{fig08}.
Time-odd terms are set to zero, so that results can be compared with
the time-even  calculations performed with {\HFBTHO} that define the
axial reference point. As apparent in Fig.~\ref{fig08}, for the
majority of configurations, triaxiality is very small, with $\gamma$
deformation typically less than 1$^{\text{o}}$.

\begin{table}[ht]
\begin{center}
\caption{Equilibrium deformation of the [402]3/2 blocked configuration
in several odd-proton isotopes with the SLy4 interaction (time-odd
terms set to zero). $\Delta^\gamma E$ represents the gain (in keV) in energy
induced by the triaxiality.}
\begin{ruledtabular}
\begin{tabular}{cccccc}
Z & N & E$^{*}$ (MeV) & $\beta$  & $\gamma$ (deg) & $\Delta^\gamma E$ (keV) \\
\hline
69 & 90 & 1.506 & 0.21 &  7.7 & -191 \\
69 & 92 & 2.070 & 0.25 &  6.7 & -187 \\
69 & 94 & 2.471 & 0.28 &  5.9 & -191 \\
69 & 96 & 2.745 & 0.29 &  5.4 & -184 \\
69 & 98 & 2.955 & 0.30 &  4.0 & -124 \\
71 & 86 & 0.232 & 0.13 & 19.6 & -233 \\
71 & 88 & 0.647 & 0.17 & 11.8 & -214 \\
71 & 90 & 1.106 & 0.20 &  8.9 & -195 \\
73 & 88 & 0.442 & 0.16 &  8.1 & -203 \\
73 & 90 & 0.717 & 0.18 &  8.9 & -205 \\
\end{tabular}
\label{table04}
\end{ruledtabular}
\end{center}
\end{table}

Only a few highly excited states are characterized by a sizeable
triaxial polarization: One such example is the state [402]3/2,
which originates from the spherical d$_{3/2}$ orbital from the
$N=4$ major shell and is pushed up into the $N = 5$ major shell
because of deformation. In Table \ref{table04} we show the
equilibrium deformations calculated with SLy4  for  this specific
configuration in a number of isotopes. The excitation energies of
[402]3/2 range from 0.2 to 3\,MeV. On average, the net energy gain
induced by the triaxial polarization of the core is of the order
of 200\,keV in this extreme case.

As indicated, the results presented in Fig.~\ref{fig08} have been
obtained by setting all time-odd fields to zero. When this constraint
is released, g.s.\ configurations remain overwhelmingly axial,
independently of the orientation of the alignment vector, cf.\ discussion
in Sec.~\ref{Sec-DFT-Symmetries}. However, we do observe that in
the collective orientation limit, low-$j$ intruder states such as
[541]1/2 (from h$_{9/2}$) and [550]1/2 (from h$_{11/2}$), or
high-$j$ intruder states such as [505]11/2 (from h$_{11/2}$), seem
slightly more unstable against $\gamma$-polarization than in
the non-collective situation.

\begin{figure}[ht]
\includegraphics[width=\columnwidth]{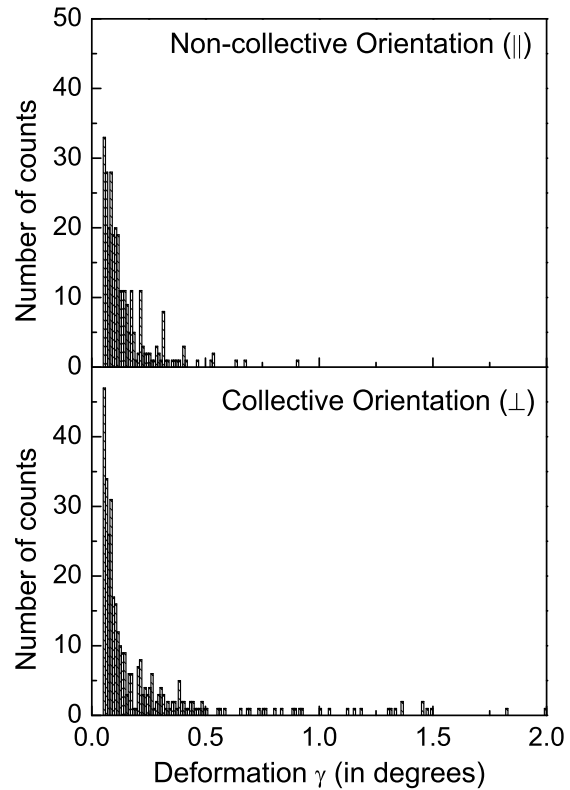}
\caption{
Triaxiality (measured by the angle $\gamma$) of well-deformed
odd-proton states ($\beta > 0.1$) in the rare-earth region
calculated with SIII, SkP, and SLy4 {\EDF}s for the two orientations
discussed in Sec.~\ref{Sec-Results-Symmetries}: collective
(upper-panel) and non-collective (lower panel).
}
\label{fig09}
\end{figure}

This overall axial stability is illustrated in Fig.~\ref{fig09}, where
the distributions of the $\gamma$ angles for well-deformed odd-proton
states in the rare-earth nuclei are plotted. For better legibility
of the figure, the very rare pronounced triaxial cases with
$\gamma > 2^{\text{o}}$ have been omitted - they have been
discussed above, and so have the many near-axial states with
$\gamma < 0.05^{\text{o}}$. In the lower panel corresponding to the
collective orientation, the few points beyond $\gamma = 1^{\text{o}}$
correspond to the $\gamma$-driving orbitals. If the rotational frequency
$\omega_{y}$ is increased (cranking), we find that the degree of
triaxiality increases accordingly \cite{[Fra83],[Ham83],[Abe90]}.


\subsection{Finite-size Instabilities of Band-head Calculations}
\label{Sec-Results-stability}

It has been shown  that some parametrizations of the Skyrme energy
functional could be prone to finite size instabilities
\cite{[Bla76],[Cau80a],[Les06]}. In particular, the time-even
$C_{t}^{\Delta\rho}\rho_{t}\Delta\rho_{t}$ and time-odd
$C_{t}^{\Delta s}\gras{s}_{t}\Delta\gras{s}_{t}$ terms could, in some
cases, lead to divergences of the {\HFB} iterative procedure. The
size of these  instabilities depends on a number of factors such as
the {\EDF} parametrization,  particle number, and specific
implementation of the {\DFT} solver. The detailed analysis of {\EDF}
instabilities performed in \cite{[Les06]} has been based on the
{\RPA} response function approach of Ref.~\cite{(Fet71)}
implemented to Skyrme functionals \cite{[Gar92b],[Mar06a]}. Results
were reported in $^{40}$Ca and $^{56}$Ni for the SkP and SLy5
parametrizations.

Finite-size instabilities governed by $C_{t}^{\Delta s}$  terms are
amplified in polarized systems such as odd-mass nuclei. Indeed, these
terms are only active when time-reversal symmetry is broken. As
was shown in Sec.~\ref{Sec-Results-Todd}, the impact of time-odd
components  is  weak, at least in the rare-earth region that we
study. It is therefore possible to scale these terms by slightly
varying the values of  $C_{t}^{\Delta s}$,  without impacting
significantly the calculated properties. By contrast, scaling the
coupling constants $C_{t}^{\Delta\rho}$ could result in
totally non-physical solutions.

According to \cite{[Les06]}, the functionals employed in this work,
namely SIII, SkP, and SLy4, should not be  particularly sensitive to
spin instabilities. Indeed, the rate of convergence in our
calculations is of the order of 40-50\% for those three cases. This
is less than for even-even axially deformed nuclei, but such a rate
can be tied to factors such as collapse of pairing,  level crossings,
etc.

\begin{figure}[ht]
\includegraphics[width=\columnwidth]{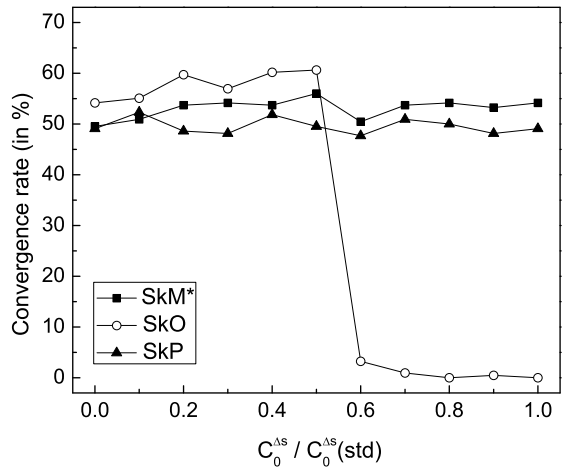}
\caption{Convergence rate of {\HFB} equations with  SkP, SkO, and SkM*
functionals for one-quasiproton states in odd-$A$ Ho isotopes with
88$\leq$$N$$\leq$104  as a function of the scalar-isoscalar coupling
constant $C_{0}^{\Delta s}$. See text for details.
}
\label{fig10}
\end{figure}

However, other Skyrme parametrizations  may be  prone to
severe and systematic divergences. To illustrate this point, we have
performed a set of calculations with three  functionals: SkO
\cite{[Rei99fw]}, SkP, and SkM* \cite{[Bar82fw]}. For each of those,
we have used the native variant of the time-odd terms; only
$C_{0}^{\Delta s}$ is multiplied by a scaling factor $\alpha$ ranging
between zero (no coupling) and one (standard coupling). A  measure of
stability of the iterative process is the rate of convergence for a
pre-defined set of one-quasiparticle states. A result is deemed
converged if the binding energy does not change by more than 2\,keV
from one iteration to the next for 3 consecutive iterations. We show
in Fig.~\ref{fig10} the evolution of this convergence rate as a
function of  $\alpha$. Our set of configurations consists of   24
different one-quasiproton states in nine odd-$A$ Ho isotopes with
88$\leq$$N$$\leq$104.  Therefore, the sample size used to define the
convergence rate is 216.

According to  Fig.~\ref{fig10}, SkM* and SkP parametrizations are
stable with respect to variations of $C_{0}^{\Delta s}$, but the
SkO functional exhibits a sharp drop in the convergence rate when $\alpha
> 0.5$, i.e., $C_{0}^{\Delta s} \gtrsim 35 $\,MeV. Preliminary
investigation of the {\RPA} response function \cite{[Les09]} suggests
that instabilities could occur for transferred momenta $q$ of the
order of 2.2--2.5\,fm$^{-1}$ for this particular value of
$C_{0}^{\Delta s}$. These results nicely agree with the original
findings of \cite{[Les06]} and emphasize the need of  testing {\EDF}s
against finite-size instabilities.


\section{Conclusions}
\label{Sec-Conclusions}

In this work, we carried out the systematic theoretical survey of
one-quasiproton states in deformed rare-earth nuclei. Our study is
based on the symmetry-unconstrained Skyrme {\HFB} framework that
fully takes into account time-odd polarization effects.

We show that the equal filling approximation is equivalent to the full
blocking when the time-odd fields are put to zero. In this case, an
arbitrary combination of time-reversed orbits can be used to define
the blocked orbit, and this can be nicely quantified by introducing
the notion of alispin. We emphasize the role of symmetries, and in
particular nuclear alignment properties, in the exact treatment of
the blocked state.

Our systematic survey indicates that, when native functionals are
employed, the contributions from time-odd fields to the energy
of the g.s.\ and low-lying excited states is rather small, around
50\,keV on average, with a variation around 100--150\,keV. Significant
differences are found from one interaction to another, although the
effect remains small for the three interactions considered.
Correcting the time-odd channel (Landau functionals) increases the
contribution of the time-odd channel to the total energy by about 50\%.
For the functionals in the gauge variant, the time-odd effects are
weak and opposite in sign.

By explicit calculations we demonstrated that the choice of the
alignment orientation with respect to the quantization axis does
impact predicted time-odd polarization energies. The resulting
energy shifts are appreciable in the scale of predicted time-odd
displacements.

Standard parameterizations of the Skyrme interaction, such as the
SIII, SkP, and SLy4, give a  qualitative, but not quantitative
description of experimental one-quasiproton spectra in the rare-earth
region. We find that the triaxial shape polarization effects are
generally small in the nuclei considered. Finally, we point to the
sensitivity of {\DFT} calculations for one-quasiparticle states to
finite-size instabilities of the underlying {\EDF}. A detailed
investigation of this effect is currently under way.

The weak impact of the time-odd fields on spectroscopic properties
implies that global studies with symmetry-restricted {\HFB} solvers
such as {\HFBTHO} could be  very useful to extract information
related to the isovector properties, shell structure, and shapes.
For such a purpose, time-odd fields may be safely neglected.

\begin{acknowledgments}
We are  thankful to T. Duguet and T. Lesinski for pointing
out finite-size instabilities as a possible explanation for the systematic
lack of convergence in odd nuclei with certain functionals. Discussions
with S.\ Fracasso are also acknowledged. This work
was supported by the U.S. Department of Energy under Contract Nos.
DE-FC02-07ER41457 (UNEDF SciDAC Collaboration), DE-FG02-96ER40963
(University of Tennessee), DE-AC05-00OR22725 with UT-Battelle, LLC
(Oak Ridge National Laboratory), and DE-FG0587ER40361 (Joint Institute
for Heavy Ion Research); by the Polish Ministry of Science and Higher
Education under Contract No. N~N~202~328234; and by the Academy of
Finland and University of Jyv\"{a}skyl\"{a} within the FIDIPRO program.
Computational resources were provided by the National Center for
Computational Sciences at Oak Ridge National Laboratory.
\end{acknowledgments}

\bibliographystyle{unsrt}

\end{document}